\documentclass[11pt,titlepage,a4paper]{article}

\usepackage{setspace} 
\usepackage{amsmath}

\usepackage{graphicx}

\usepackage{hyphenat}
\usepackage{placeins}
\usepackage{changepage}
\usepackage[singlelinecheck=false,font=small,labelfont=bf]{caption}
\usepackage{color,soul}

\usepackage[round,numbers,sort&compress]{natbib} 

\newcommand{\ATPfrac}{\ensuremath{\alpha_{\mathrm{ATP}}}}
\newcommand{\npT}{\ensuremath{n_{\mathrm{T}}}}
\newcommand{\npS}{\ensuremath{n_{\mathrm{S}}}}

\newcommand{\rmlabels}[3]{\ensuremath{#1^{\mathrm{#2}}_{\mathrm{#3}}}}

\def\longrightharpoonup{\relbar\joinrel\rightharpoonup}
\def\longleftharpoondown{\leftharpoondown\joinrel\relbar}

\def\longrightleftharpoons{
  \mathop{
    \vcenter{
      \hbox{
      \ooalign{
        \raise1pt\hbox{$\longrightharpoonup\joinrel$}\crcr
	  \lower1pt\hbox{$\longleftharpoondown\joinrel$}
	  }
      }
    }
  }
}

\newcommand{\rates}[2]{\displaystyle
  \mathrel{\longrightleftharpoons^{#1\mathstrut}_{#2}}}

\newcommand{\tref}[1]{Table~\ref{table:#1}}

\newcommand{\tlabel}[1]{\label{table:#1}}

\newcommand{\fref}[1]{Fig.~\ref{fig:#1}}

\newcommand{\flabel}[1]{\label{fig:#1}}

\newcommand{\eref}[1]{Eq.~\ref{eqn:#1}}

\newcommand{\elabel}[1]{\label{eqn:#1}}

\title{Period robustness and entrainability under changing nucleotide concentrations in the post\hyp{}translational Kai circadian clock}

\author{Joris~Paijmans\\
	AMOLF, \\
	Science Park 102, 1098 XG Amsterdam, The Netherlands
	\and David K. Lubensky \\
	Department of Physics, \\
	University of Michigan, Ann Arbor, MI 48109-1040, USA
	\and Pieter Rein ten Wolde\thanks{
           Corresponding author.  Address: 
	   AMOLF,
	   Science Park 102,
	   1098XG, Amsterdam, The Netherlands,
	   email: tenwolde@amolf.nl} \\
	AMOLF, \\
	Science Park 102, 1098 XG Amsterdam, The Netherlands
	}

\date{}

\begin{document}

\maketitle

\abstract{Circadian clocks must be able to entrain to time-varying signals to keep their oscillations in
  phase with the day-night rhythm. On the other hand, they must also
  exhibit input compensation: their period must remain about one day in different
  constant environments.
  The post-translational oscillator
  of the Kai system can be entrained by transient or oscillatory
  changes in the ATP fraction, yet is insensitive to constant changes
  in this fraction. We study in three different models of this system
  how these two seemingly conflicting criteria are met: 
  the Van Zon model (Van Zon et al., PNAS, 2007), the Rust model (Phong et al., PNAS, 2013), 
  and the Paijmans model (Paijmans et al., PLoS Comput. Biol., 2017). 
  We find that the Paijmans model exhibits the best trade-off between input compensation 
  and entrainability: on the footing of equal phase-response curves, it
  exhibits the strongest input compensation. Performing stochastic
  simulations at the level of individual hexamers allows us to
  identify, to the best of our knowledge, a new mechanism, which is employed by the Paijmans model to
  achieve input compensation: At lower ATP fraction, the individual
  hexamers make a shorter cycle in the phosphorylation state space,
  which compensates for the slower pace at which they traverse the cycle.\\
  
}

\clearpage

\section*{Introduction}
Circadian clocks help organisms to
coordinate their metabolism and behavior with the daily changes in the
environment \cite{Winfree1980}. These clocks are prevalent in a wide
range of organisms from bacteria to humans, but all have three
important features in common: First, a circadian clock is a
\emph{self\hyp{}sustained oscillator}, meaning that oscillations
persist even in the absence of any external cue, with a rhythm of about 24 hours.  
Second, the clock is \emph{entrainable} so that its oscillations can be kept in
phase with the environments day-night rhythm.
To this end, a circadian clock must
be able to respond to daily cues such as rhythmic changes in light and temperature. 
Third, the clock has some form of \emph{input compensation}, 
such that the period is constant, even when the
temperature or light intensities change for longer times.
Importantly, these last two requirements seem to be at odds with each other. 
A clock that is easily entrained, because, for instance, the
rates of the biochemical reactions that make up the clock strongly
depend on temperature, would normally be expected to have a period that also depends on the temperature. 
This clock would be a bad predictor of time.
As was shown in \cite{Ouyang1998,Woelfle2004}, 
clock mutants having intrinsic periods ranging from 22 to 30 hours,
are outcompeted in a culture with wild-type cells with a period closer 
to 24 hours when growing under a 12:12 light-dark cycle.
Therefore, input compensation is critically important to the function of a circadian clock.
On the other hand, if a clock would be completely
insensitive to any external change, it would trivially have a period
that is insensitive to changes in the environment. 
However, such a clock would not be entrainable, and as a result of biochemical
noise, it would inevitably run out of phase with the day-night rhythm. 
What the properties of an oscillator should be to fulfill both conditions 
has been studied in different systems, both experimentally and theoretically
\cite{Lakin-Thomas1991,Brettschneider2010,Pfeuty2011,Hatakeyama2012,Hatakeyama2015,Hasegawa2014,Kidd2015}.

We use the post\hyp{}translational oscillator of the Kai circadian clock, 
found in the freshwater cyanobacterium \emph{Synechococcus elongatus} PCC 7942,
as a model system.
It is well known that this circadian clock is sensitive to transient changes 
in its environment,
and hence entrainable, yet robust to permanent changes in its input,
thus showing input compensation \cite{Kondo1993, Schmitz2000, Kitayama2008, Qin2010, Hosokawa2013}. 
It was shown in a seminal experiment in 2005, 
that the core oscillator can be reconstituted in-vitro, 
and consists of the three proteins KaiA, KaiB and KaiC, in solution with ATP \cite{Nakajima2005}.  
KaiC is a phosphotransferase \cite{Nishiwaki2012, Egli2012}, 
which, depending on its conformation,
switches between phases of auto-phosphorylation and auto-dephosphorylation
\cite{VanZon2007, Kim2008}.  KaiA is a nucleotide exchange factor,
that facilitates exchange of ADP to ATP in the nucleotide binding
pockets of KaiC \cite{Nishiwaki-Ohkawa2014}, which enhances
phosphorylation of KaiC. 
KaiB counteracts the effect of KaiA, 
by binding to KaiC, and sequestering KaiA from solution \cite{Williams2002, Xu2003}.

Remarkably, even in an in-vitro essay, the phase of the oscillator can be reset while keeping a
robust circadian period, by changing the temperature \cite{Nakajima2005, Yoshida2009, Murayama2011} 
or the ratio of ATP to ADP in the buffer \cite{Rust2011, Phong2012}.
Also the redox state can change the phase of the oscillator \cite{Wood2010, Kim2012},
but the period is not maintained as the oscillations are disrupted under oxidized conditions.
In this work, we study entrainability and input compensation
under changes in the bulk ATP fraction, by comparing three different
models of the post\hyp{}translational Kai oscillator. 
It is known that the ATP fraction in the cytoplasm of \textit{S. elongatus}, 
due to photosynthesis, fluctuates between 90\% during the day down to 30\% at night \cite{Rust2011}.
Experiments show that a dark pulse induces a gradual drop in the cell's ATP level at any time of the day,
which in turn induces a phase shift of the circadian clock \cite{Rust2011}.
To confirm that the ATP level is directly responsible for this shift, 
Pattanayak \textit{et. al.} constructed a transgenic cyanobacterium which produces an enzyme that converts glucose into ATP, 
and thus maintains a steady ATP level under a light-dark cycle when supplied with glucose \cite{Pattanayak2015}. 
It turns out that the Kai oscillator in this transgenic cyanobacterium is no longer entrained by light-dark pulses. 
From this the authors conclude that metabolic activity, 
independent of its source, is the primary Zeitgeber in \textit{S. elongatus}.
We therefore chose to investigate the influence of the ATP fraction on the Kai oscillator.

Specifically, we will study the effect of the ATP fraction on the hexamer model by Van Zon \textit{et. al.} \cite{VanZon2007}; 
the extended monomer model, introduced by Phong and coworkers in \cite{Phong2012} as an extension of the simpler model
originally proposed in \cite{Rust2007}; and the model introduced in \cite{Paijmans2017}
which we refer to as the ''Paijmans model'' and which accounts for both KaiC its 
hexameric form and the fact that each KaiC monomer contains two distinct phosphorylation sites.
In general, as shown in experiments, the effect of the ATP fraction on the oscillator
is that, during the phosphorylation phase of the oscillation, 
the phosphorylation rate is roughly proportional to the ATP fraction,
while during the dephosphorylation phase, the overall rate of dephosphorylation is unaffected \cite{Rust2011}. 
We will study the entrainability by applying a 6 hour pulse, 
during which the ATP fraction is reduced from 100\% to 40\%, 
at different phases of the oscillation, 
and compare the maximally induced phase shift in each model. 
We analyze period stability in each model by running the
models at different constant ATP fractions and observe how the period
and other important quantities of the oscillations change.  Lastly,
simulations of our Paijmans model allow us to track each individual
hexamer, and measure the timing between states as it proceeds through
its cycle.  In this way, to our knowledge for the first time, we can
predict how individual KaiC hexamers respond to external cues.

Below, the theory section briefly describes the three models, and the mechanisms they
employ to achieve input compensation and entrainability.  
We discuss three mechanisms of input compensation, 
two of which have been partly identified elsewhere, and one of which is novel.
The first, present in all models, is related to the delay between the moment the
KaiC front runners (hexamers that are more phosphorylated than the average) 
reach the top of their phosphorylation cycle, 
which is the point they no longer need KaiA to progress along the cycle, 
and the later time at which the front runners reach a state 
in which they sequester KaiA \cite{Phong2012}. 
As the ATP fraction in the bulk is reduced, the
rate of phosphorylation decreases, which means that it takes longer to
reach the top of the cycle, thus extending the phosphorylation
phase. However, the lower rate of phosphorylation also means that
during the delay (whose length is independent of the ATP fraction) 
less hexamers make it to the top of the cycle. 
This decreases the number of KaiC molecules that can participate in
sequestering KaiA, thus shortening the dephosphorylation phase, 
counteracting the longer phosphorylation phase.  The second mechanism,
present only in the Rust model, is related to the positive feedback in
that model which results from the mutual inhibition between the
sequestration of KaiA by serine-phosphorylated KaiC, 
and KaiA stimulating the transition from serine-phosphorylated to doubly phosphorylated KaiC.
The third mechanism, present in our Paijmans model, and first identified here, 
concerns the path individual KaiC hexamers take through phosphorylation state space: 
At lower ATP fraction, the individual hexamers move through a smaller phosphorylation cycle, 
which compensates for the lower rate at which they traverse this cycle.
The generic idea of input compensation via a trade-off between the size of the cycle in state space and the speed 
at which it is traversed was presented by Lakin-Thomas \textit{et. al.}\cite{Lakin-Thomas1991} and later in more detail by Hatakeyama and Kaneko \cite{Hatakeyama2012,Hatakeyama2015}. 
Here we present a specific manifestation of this mechanism at the level of individual hexamers.
This mechanism is absent in the Van Zon and Rust models, 
where the phosphorylation cycles of the hexamers or monomers 
through state space are independent of the ATP fraction.

As period robustness can trivially be achieved by making the
oscillator completely insensitive to the ATP fraction, the results section
investigates the phase response curve for each model, which describes
the induced phase shift of the clock upon a pulse of ADP
\cite{Winfree1980}.  We find that the models of Van Zon and Rust are,
depending on their sensitivity to the ATP fraction, either strongly
entrainable yet have a period that depends on the ATP fraction or have
a very stable period but are not entrainable. In contrast, our Paijmans
model exhibits both input compensation and entrainability.

\section*{Theory}
In this section we give a description of the three models
of the post\hyp{}translational Kai circadian clock studied here and of
how we adapted each model to include the sensitivity of the phosphorylation rates to the ATP fractions in the bulk. 
The only model parameter other than ATP fraction that we will vary in these models is the relative affinity for ATP and ADP.
Because the other model parameters, such as the slow phosphorylation rates, 
are often strongly constrained by experimental results we keep them fixed and use the published values.
We explain which mechanisms for period stability are present in each model, 
to compensate for the dependence of the phosphorylation rates on the ATP fraction. 
Two of the three mechanisms that we describe below have been partly
presented elsewhere \cite{Phong2012}, yet we will discuss here how they are implemented
specifically in the respective models.  For completeness, we give
here also a qualitative description of the third, novel, mechanism, which is
employed only by the Paijmans model; this mechanism is discussed in much
more detail in the Results section.
We first give an overview of the most important features of the Kai system.

The principal pacemaker of the Kai circadian clock consists of three proteins, KaiA, KaiB and KaiC \cite{Akiyama2012}.
The only enzymatic component is KaiC, which forms a homohexamer, and consists of two domains, CI and CII.
Both domains can bind and hydrolyze ATP, but only the CII domain can be phosphorylated.
The CII domain has two phosphorylation sites, the threonine and the serine site, 
resulting in four different phosphorylation states for each monomer \cite{Rust2007, Nishiwaki2007}:
unphosphorylated (U), phosphorylated only on serine (S),
phosphorylated only on threonine (T), and double phosphorylated on both
serine and threonine, (D).
Because the phosphorylation rate of the threonine site is much higher than for the serine site,
and because the binding of KaiA to the CII domain of KaiC suppresses the phosphorylation
of the serine site, the threonine site is usually phosphorylated before the serine site.
The result is that in each oscillation a KaiC monomer typically goes through an ordered phosphorylation cycle: U,T,D and S.
The hexamer can be in two conformations which we call the active and inactive states, 
and the phosphorylation state will determine which conformation the hexamer is in.
In the active state, KaiA can bind to the CII domain and promotes the phosphorylation of KaiC
and in the inactive state, KaiB can bind to the CI domain, which will then sequester and thereby inactive KaiA,
such that all KaiC will dephosphorylate.

\subsection*{Van Zon model}
The Van Zon model describes the phosphorylation cycle at the level of
KaiC hexamers, and does not explicitly keep track of the KaiC monomers \cite{VanZon2007}. 
A simplified scheme of the model is shown in
\fref{StabilityMechanisms}A: A hexamer can be in the active
conformational state, denoted by C$_i$, or in the inactive state,
denoted by $\tilde{\mathrm{C}}_i$, where $i$ denotes its phosphorylation level. 
In the presence of free KaiA, active KaiC
(enclosed by the green box in \fref{StabilityMechanisms}A) is
phosphorylated with a rate which dependents on the KaiA concentration.
When the hexamer has reached the fully phosphorylated state,
$\mathrm{C}_{6}$, it flips to the inactive conformation,
$\mathrm{C}_{6}\rightarrow\mathrm{\tilde{C}}_{6}$, where it immediately binds KaiB. 
The delay between full phosphorylation and KaiA sequestration, 
essential for synchronized oscillations, is set by the two dephosphorylation steps,
$\mathrm{\tilde{C}}_{6}\rightarrow\mathrm{\tilde{C}}_{5}\rightarrow\mathrm{\tilde{C}}_{4}$
(blue box).  The complexes
$\mathrm{\tilde{C}}_{4}-\mathrm{\tilde{C}}_{1}$, have a very high
affinity for KaiA, allowing them to sequester all free KaiA from the
solution (red box).  This sequestration of KaiA forces the front
runners that have reached the bottom of the cycle ($\mathrm{C}_{0}$)
and are ready to be phosphorylated again, to wait, because KaiA is needed for
phosphorylation.  
Sequestration of KaiA thus allows the laggards 
(the hexamers that are falling behind in the phosphorylation cycle)
to catch up with the front runners, 
leading to the synchronization of the oscillations of the individual hexamers. 
Only when most KaiC has reached $\mathrm{C}_0$, is KaiA released in solution, 
and can a new phosphorylation cycle start again.

We use a coarse grained description to model the effect of the ATP
fraction, \ATPfrac{}=[ATP]/([ATP]+[ADP]), where [ATP] and [ADP] are
the ATP and ADP concentrations in the bulk, respectively, on the
phosphorylation rates.  Just like in the Rust model \cite{Rust2011},
we assume that, when KaiA is bound, the probability of having ATP
instead of ADP bound, is
\begin{equation}
 \elabel{betaATP}
 \rmlabels{\beta}{}{ATP}=\frac{\ATPfrac{}}{\ATPfrac{}+\rmlabels{K}{}{ATP/ADP}(1-\ATPfrac{})},
\end{equation}
where \rmlabels{K}{}{ATP/ADP} is the relative dissociation constant for binding ATP over ADP.
The effective phosphorylation rates become $\rmlabels{k}{}{phos}=\rmlabels{\beta}{}{ATP}\,\rmlabels{k}{0}{phos}$,
where \rmlabels{k}{0}{phos} is the phosphorylation rate at 100\% ATP.
Dephosphorylation rates are independent of \ATPfrac{}.

The Van Zon model employs one mechanism for period stability,
which is partly identified in \cite{Phong2012}.
The mechanism is a direct consequence of the temporal delay between the moment a hexamer 
reaches the top of the cycle, i.e. the state $\mathrm{\tilde{C}}_6$
in which it no longer needs KaiA to progress along the cycle, 
and the time at which it reaches $\mathrm{\tilde{C}}_4$ and starts sequestering KaiA. 
Thus, in particular, there is a lag between the moment when enough KaiC hexamers 
to fully sequester KaiA have passed $\mathrm{\tilde{C}}_6$, 
and so are committed to the path towards sequestration, 
and the moment when full sequestration is actually reached. 
The number of additional hexamers that reach $\mathrm{\tilde{C}}_6$
during this delay is given by the duration of 
the delay multiplied by the phosphorylation speed. 
Importantly, while the duration of the delay is independent of the ATP fraction 
(since dephosphorylation from $\mathrm{\tilde{C}}_6$ till $\mathrm{\tilde{C}}_4$ is independent of the ATP fraction),
the rate of phosphorylation decreases as the ATP fraction decreases. 
Consequently, the lower the ATP fraction, 
the smaller the number of hexamers that can reach the state $\mathrm{\tilde{C}}_6$ during the delay. 
The smaller number of $\mathrm{\tilde{C}}_6$, in turn, 
leads to a shorter time interval in which all KaiA stays sequestered. 
This shortens the dephosphorylation phase,
which counteracts the longer phosphorylation phase, stabilizing the period.

\fref{StabilityMechanisms}B shows time traces for the KaiC
phosphorylation level
$p(t)=\sum_{i=1}^{6}i(\mathrm{C}_{i}+\mathrm{\tilde{C}}_{i})/(6\,\mathrm{KaiC_{tot}})$
(dotted lines), and the fraction of inactive hexamers (solid lines),
at \ATPfrac{}=100\% and 50\%.  
The number of hexamers that can sequester KaiA increases with 
the amplitude of the inactive fraction,
such that this amplitude sets the duration of the dephosphorylation phase.
Indeed, as a clear signature of the stability mechanism, at 50\% ATP fraction,
both the phosphorylation level and the fraction of inactive KaiC rise
slower while having a lower amplitude compared to the oscillations at
100\% ATP fraction.

\subsection*{Rust model}
Contrary to the Paijmans model and the Van Zon model, 
the Rust model describes the oscillations at the level of single monomers \cite{Rust2007, Rust2011, Phong2012}.
As shown in \fref{StabilityMechanisms}C, each monomer goes through the ordered phosphorylation cycle
$\mathrm{U}\rightarrow\mathrm{T}\rightarrow\mathrm{D}\rightarrow\mathrm{S}\rightarrow\mathrm{U}$.
Only when mono\-mers have reached the phosphorylation states S and D, they can bind KaiB with a low rate,
and form D$\cdot$B and S$\cdot$B, respectively.
Importantly, only the S$\cdot$B state sequesters KaiA,
while KaiA impedes the occupation of the S$\cdot$B state by enhancing the transition
from S$\cdot$B back to D$\cdot$B.
This mutual inhibition between KaiA and the S$\cdot$B state creates a positive
feedback loop for KaiA sequestration that is essential for oscillations.
Initially, when D$\cdot$B transforms into S$\cdot$B,
KaiA stimulates the reverse reaction.
During this period of a quasi-equilibrium  between the D$\cdot$B and S$\cdot$B states,
the concentration of their sum rises, [D$\cdot$B] + [S$\cdot$B], 
up to the point that [S$\cdot$B] reaches a level where it sequesters all KaiA.
At this moment, the positive feedback is broken, 
and the system rapidly switches to the dephosphorylation phase in which KaiA is fully sequestered for a long time.
The positive feedback thus creates a sharp transition 
between the phase in which KaiA is free to simulate phosphorylation,
and the phase in which all KaiA is sequestered.

To include the effect of the bulk ATP fraction, 
we use the same coarse grained description on the phosphorylation rates as in the original work: 
$\rmlabels{k}{}{phos}=\rmlabels{\beta}{}{ATP}\,\rmlabels{k}{0}{phos}$,
where \rmlabels{\beta}{}{ATP} is defined in \eref{betaATP}.
As was shown in \cite{Phong2012}, 
the ATPase activity in the CI domain, 
which sets the rate of KaiB binding in this model,
does not depend on the bulk ATP fraction.

The Rust model implements two methods for period stability,
where the first is similar to that identified in the Van Zon model and described by Phong \textit{et. al.} \cite{Phong2012}.
The slow KaiB binding step creates a temporal delay between, on the one hand, 
the phosphorylation states D and S,
which can only be reached in the presence of free KaiA,
and, on the other hand, the state S$\cdot$B, which sequesters KaiA.
This delay allows monomers to reach the D and S state through phosphorylation before all KaiA is sequestered.
The number of monomers that can reach the KaiB bound states increases with the phosphorylation speed,
set by $\ATPfrac{}$, 
and will determine the duration of the period in which all KaiA is sequestered.

The previous mechanism assumes that the delay between reaching the D or S phosphorylated states 
and sequestering KaiA is constant, 
but the input compensation can be enhanced further if this delay gets shorter when ADP is added.
More precisely, in order to sequester all KaiA, the concentration of S$\cdot$B monomers has to fulfill
$\rmlabels{n}{S}{seq}[\rmlabels{(S\cdot B)}{}{min}]\geq[\mathrm{KaiA}_{\mathrm{tot}}]$,
where \rmlabels{n}{S}{seq} is the number of KaiA monomers sequestered 
by a single S$\cdot$B KaiC monomer and $[\rmlabels{(S\cdot B)}{}{min}]$ is the minimal 
concentration of S$\cdot$B monomers to sequester all KaiA, $\mathrm{KaiA}_{\mathrm{tot}}$.
Because the transitions between the states S$\cdot$B and D$\cdot$B 
are faster compared to the transitions from these states to the U and T,
the S$\cdot$B$\rates{}{}$D$\cdot$B transitions are in quasi equilibrium.
This means that [D$\cdot$B] is related to [S$\cdot$B] via the steady state relation,
[D$\cdot$B]/[S$\cdot$B]$\approx\rmlabels{k}{}{SD}(\ATPfrac{})/\rmlabels{k}{}{DS}=f(\ATPfrac{})$,
which depends on the effective (de)phosphorylation rates, 
\rmlabels{k}{}{SD}(\ATPfrac{}) and \rmlabels{k}{}{DS}, respectively,
and, importantly, on the bulk ATP fraction.
Consequently, the amount of D$\cdot$B required to have enough S$\cdot$B to sequester all KaiA,
is D$\cdot$B$\approx$S$\cdot$B$\,f(\ATPfrac{})$, which is related to \ATPfrac{}.
Thus, at a lower bulk ATP fraction, 
the concentration of KaiC bound to KaiB necessary to sequester all KaiA, [D$\cdot$B]+[S$\cdot$B], 
will be lower, which compensates for the slower formation of these complexes during the phosphorylation phase.
Indeed, as shown in \fref{StabilityMechanisms}D, the concentration of KaiC-bound KaiB 
increases much slower at a 50\% ATP fraction compared to 100\%,
but the concentration of KaiC-bound KaiB at the moment that all KaiA is sequestered, 
is also lower.
Because fewer D phosphorylated monomers are required to sequester all KaiA,
the subsequent sequestration time is smaller, shortening the period even more.

\subsection*{Paijmans model}
The Paijmans model, explained in more detail in \cite{Paijmans2017}, is again a hexamer model.
This model, shown in \fref{StabilityMechanisms}E, explicitly describes the state of individual monomers,
and in particular their serine and threonine phosphorylation sites.
Each monomer in a hexamer is phosphorylated in a well defined order: 
First the threonine site is phosphorylated and then the serine site.
Phosphorylation of the two sites has an antagonistic effect on the conformational state of the hexamer: 
The U and T states stabilize the active conformation and the D and S states stabilize the inactive conformation.
Due to this antagonism, the relative stability of the conformations do not depend 
on the absolute number of monomers in a certain state, 
but rater on the difference between the number of phosphorylated threonine and serine sites \cite{Lin2014}.
Roughly, when more serine sites are phosphorylated than threonine sites,
the hexamer will switch conformation.
After flipping to the inactive state, the hexamer binds KaiB, 
but it can only sequester KaiA after 6 KaiB monomers are bound.
This delay allows hexamers lagging behind to continue phosphorylation and reach the inactive state,
which is an essential property of our model to generate robust oscillations.

The Paijmans model explicitly simulates the binding and unbinding of nucleotides,
and the hydrolysis of ATP, in the CII domain,
where the ATP is used to phosphorylate the threonine and serine sites.
The ATP fraction in the CII binding pocket is dependent on the bulk ATP fraction,
because the binding rate of nucleotides is directly proportional to \ATPfrac{}.
Furthermore, the ATP fraction in the binding pocket depends on the hydrolysis 
rate of ATP, and the relative affinity for ATP and ADP.
Importantly, both the effective phosphorylation \emph{and} dephosphorylation rates depend 
on the ATP fraction of the binding pockets, 
because both events occur via phosphotransfer with the bound nucleotide.
This means that a change in \ATPfrac{} has a much bigger effect on the (de)phosphorylation dynamics
compared to models where dephosphorylation proceeds through a \ATPfrac{} independent phosphatase reaction.

We choose the relative affinity of nucleotides for the binding pockets of the CII domain, $\rmlabels{K}{CII}{ATP/ADP}=0.1$, 
much lower than what is used in the Rust model \cite{Rust2011}.
As shown in \cite{Paijmans2017}, this value gives a good agreement with experiments which measure 
the sensitivity of the phosphorylation dynamics on the bulk ATP fraction 
in a system starting with KaiA and unphosphorylated KaiC only \cite{Rust2011, Phong2012}.
We have, however, also investigated the behavior of the other two models
with a relative affinity that is similar as used in the Paijmans model.
In the results section below, 
we will compare the other models with both high and low relative affinity to the Paijmans model.

The oscillator employs two mechanisms of period stability.  First, due
to the slow KaiB binding, there is a delay between flipping to the
inactive conformation after phosphorylation, and KaiA sequestration.
This creates a pool of hexamers in the inactive conformation that 
increases with the speed of phosphorylation, set by the ATP fraction.
Again, at lower \ATPfrac{}, fewer hexamers make it to the inactive
state, such that the dephosphorylation phase is shorter, which
counteracts the longer phosphorylation phase.  

The second mechanism for period stability is related to the path individual hexamers traverse through 
phosphorylation state space before switching to the inactive conformation.
A hexamer switches to the inactive state when the number of phosphorylated serine sites, 
\npS{}{}, exceeds the number of phosphorylated threonine sites, \npT{}.
Due to the ordered phosphorylation of each monomer, the threonine sites are phosphorylated before the serine sites,
such that a hexamer makes a wide arc in phosphorylation state space, (\npT{},\npS{}{}),
before the diagonal, and the flipping criterion $\npS{} > \npT{}$, is reached.
As the effective phosphorylation rates decrease for lower \ATPfrac{}, 
the size of the arc  decreases, as discussed in more detail below.
This shorter path in state space counteracts the effect of a slower
progression along the path (due to the slower phosphorylation),
and creates another mechanism for input compensation, at the level of inidividual hexamers.

Note that the positive feedback loop on KaiA sequestration in the Rust
model, due to the mutual repression between KaiA and the S$\cdot$B
state of KaiC, is not present in the Paijmans model.  In this feedback loop in
the Rust model, KaiA stimulates the transition from the S to the D
state, thereby preventing its own sequestration (because only the  S$\cdot$B
state significantly sequesters KaiA).  However, because in
the Paijmans model both the S and the D state stabilize the ADP bound state in
the binding pocket of the CI domain, which in turn stabilizes the inactive
conformation and the subsequent KaiB binding (leading to KaiA sequestration), KaiA does not prevent
its own sequestration by stimulating the S to D transition.
Therefore, the mechanism of input compensation in the Rust model
resulting from the positive feedback loop, does not apply to the Paijmans model.

\section*{Results}
\subsection*{Dependence of oscillations on the ATP fraction reveals input compensation on the ensemble level}
To find out how effective the mechanisms for period stability are
in the three models, we run simulations of the models at constant bulk
ATP fractions from 100\% to 50\%.  For the Van Zon and Rust model, we
consider two values of the relative dissociation constant for ATP
versus ADP: 1) With equal affinity for ATP and ADP,
$\rmlabels{K}{}{ATP/ADP}=1.0$ (solid lines in
\fref{ModelMeanATPDependence}), as used and motivated in the original
model by Rust \cite{Rust2011}, and 2) With a lower affinity for ADP,
$\rmlabels{K}{}{ATP/ADP}=0.19$ (dashed lines), such that the effect on
the ATP fraction in nucleotide binding pocket, \rmlabels{\beta}{}{ATP}
(\eref{betaATP}), as \ATPfrac{} decreases from 100\% to 50\%, is
similar to the drop in the Paijmans model of about 15\% (see Methods section).
As we motivated above, the relative affinity in the Paijmans model is $\rmlabels{K}{}{ATP/ADP}=0.1$.

\fref{ModelMeanATPDependence}A shows how the period varies with
decreasing \ATPfrac{}.  Remarkably, the three models have a different
response to lowering \ATPfrac{}: Whereas in the Paijmans model the period
is almost constant, in the Van Zon model it increases by 20\%, while
in the Rust model it decreases by 20\%, as \ATPfrac{} decreases from
100\% to 50\%.  This is reflected in the change in the amplitude of
the phosphorylation levels, panel B, which decreases the strongest in
the Rust model and the least in the Van Zon model, with decreasing
\ATPfrac{}.  The results in panels A and B are consistent: given the
period stabilization mechanism due to the delay, we expect that in all
models fewer hexamers or monomers make it through the cycle as
\ATPfrac{} decreases, such that the amplitude of the oscillation
decreases, shortening the length of the dephosphorylation phase.  This
view is further supported in panels C and D: Panel C shows the length
of the phosphorylation phase, $\Delta\rmlabels{t}{}{phosphorylation}$,
defined as the time between a trough and the next peak in
phosphorylation level.  As expected, this time increases in all models
as \ATPfrac{} decreases: due to the lowering of the phosphorylation
rates, it takes more time to reach the required phosphorylation state
to sequester enough KaiA.  The length of the subsequent
dephosphorylation phase, $\Delta\rmlabels{t}{}{dephosphorylation}$,
defined as the time between a peak and the next trough in the
phosphorylation level, decreases in the Paijmans model and the Rust model,
as a result of the stability mechanism.  However, in the Van Zon
model, the stability mechanism does not work as
$\Delta\rmlabels{t}{}{dephosphorylation}$ increases with decreasing \ATPfrac{}.  
Because KaiC is unable to sequester all KaiA in the system during the dephosphorylation phase,
the phosphorylation of active hexamers continues during this phase, 
which decreases the {\em net} dephosphorylation rate and extends its duration.

We point out that as the phosphorylation rates decrease,
the number of hexamers and monomers that traverse a full cycle each period, 
and sequester KaiA, decreases.
To show the effect of \ATPfrac{} on this fraction,
we plot the total flux of monomers or hexamers that move through the cycle per period, 
described in more detail in \cite{Paijmans2017} and in the Methods section, 
for each model in \fref{ModelMeanATPDependence}E.
Panel E shows a decrease in flux of around 15\% for the Van Zon and the Paijmans model at $\ATPfrac{}=50\%$.
The Rust model has a much larger decrease in flux of around 60\%, 
which confirms the idea that input compensation is achieved in this model by letting fewer monomers 
participate in the cycle and sequester KaiA per period.
Panel F gives the time interval per period when all KaiA is sequestered by KaiC, 
$\Delta\rmlabels{t}{}{KaiA\,sequestered}$, as defined in the methods section,
which shows that in all models the time of full sequestration indeed shortens (see Methods section).
Here, remarkably, the decreases upon lowering \ATPfrac{} is the strongest for the Van Zon model,
even though the decrease in the fraction of hexamers that go through a cycle and sequester KaiA (panel E)
is much less.
This is probably related to the fact that in the Van Zon model a hexamer can only sequester 2 KaiA dimers,
while in the Paijmans model 6 dimers per hexamer and in the Rust model 2.5 dimers per monomer are sequestered.
Therefore, the number of inactive hexamers required to sequester all KaiA, 
is much higher in the Van Zon model, compared to the other two models.
Hence, a small change in the amplitude in the concentration of inactive hexamers,
has a big effect on $\Delta\rmlabels{t}{}{KaiA\,sequestered}$. 
Note that below $\ATPfrac{}$ of 65\%, KaiA is never fully sequestered in the Van Zon model, 
while oscillations persist.

Comparing results between different relative affinities, \rmlabels{K}{}{ATP/ADP},
it is clear that both the Van Zon and the Rust model are much less affected by the bulk ATP fraction,
when we choose a lower relative affinity for ADP, \rmlabels{K}{}{ATP/ADP}=0.19.
For lower \rmlabels{K}{}{ATP/ADP}, the probability that the binding pocket of the CII domain 
is bound to ATP instead of ADP, is much less affected by changes in \ATPfrac{}.
This makes the phosphorylation rates, and hence the oscillations, less sensitive to \ATPfrac{}.
Note that insensitivity of the period to \ATPfrac{} can trivially be achieved in all three models 
by choosing a very low relative affinity for ADP, $\rmlabels{K}{}{ATP/ADP}\to 0$.
A robust period, however, is not enough for a biological clock to be a good predictor of time.
The key point is that a good biological clock has a period which is insensitive to the average level of \ATPfrac{},
while, to keep it in phase with the day in the presence of the inevitable biochemical noise,
it is still entrainable by periodic variations in this quantity.
Therefore, in the next section, we study how each model of the Kai oscillator 
responds to a transient lowering of the ATP fraction.

\subsection*{Van Zon and Rust models only show strong entrainability at equal nucleotide affinities}
We expose the oscillator, running at a 100\% ATP fraction, to a
transient, six hour pulse of a 40\% ATP fraction, 
starting at different times from the last trough in the phosphorylation fraction.
To calculate the phase shift induced,
we compare the time trace of the phosphorylation fraction two troughs after the onset of the pulse,
with a control where no pulse is given (see Methods section).

\fref{ModelATPSensitivity}, panels A, B and C, show phase response
curves for the Paijmans model, the Van Zon model and the Rust model,
respectively, showing the induced phase advances.  Comparing the Paijmans
model, with $\rmlabels{K}{}{ATP/ADP}=0.1$, with the other two models
with equal nucleotide affinities, $\rmlabels{K}{}{ATP/ADP}=1.0$ (solid
lines), we see that the phase response curves are comparable: Both the
maximally induced phase advance and delays are around 4 hours in all
models, although the Rust model is capable of a particularly large
phase advance of almost 10 hours. 
All models have a dead zone, where the pulse does not induce a phase change, 
starting about 12 hours from the trough, which is thought to be essential for entrainment
\cite{Pfeuty2011}. 
However, when we reduce in the Van Zon and Rust models the relative affinity 
to $\rmlabels{K}{}{ATP/ADP}=0.19$ (dashed lines), 
the amplitude of the phase response curves become much smaller: 
In the Van Zon model the maximal phase advance and delay
are reduced to around one hour. In the Rust model, the maximal phase
delay is also reduced to about one hour, which is significantly
smaller than that in the Paijmans model; the maximal phase advance is
also strongly reduced, although it is reduced to a value that is
only slightly lower than that in the Paijmans model. 
The phase response curve of the Rust model is thus asymmetric:
an ADP pulse can induce a much stronger phase advance than a phase delay.
However, it was experimentally shown that the phase response curve of the in-vitro Kai system 
is very symmetric \cite{Rust2011,Pattanayak2015}. 
Therefore, comparing \fref{ModelATPSensitivity}A-C with \fref{ModelMeanATPDependence}A
shows that the Paijmans model has the strongest similarity with the experimental data 
while it also gives the best trade-off between entrainability and input compensation: 
On the footing of equal phase response curves, \fref{ModelATPSensitivity}A-C, 
the Paijmans model has the smallest change in the period upon changing \ATPfrac{}, 
\fref{ModelMeanATPDependence}A.
We also note here that the symmetry of the PRC of the Paijmans model could be important, 
because theoretical analysis indicates that the ability to accurately entrain to external signals 
strongly depends on the shape of the PRC \cite{Pfeuty2011}.

We want to know how the phase advance and delay is achieved in each model.
To this end, in panels D-I, 
we look at time traces of the phosphorylation level (solid black lines) 
and the fraction of sequestered KaiA (solid purple lines),
when a pulse of ADP is given (shaded regions),
where we set $\rmlabels{K}{}{ATP/ADP}=1.0$ for the Van Zon and Rust models.
For comparison, the dashed lines show the control where no pulse is given.
We choose the starting times of the pulses such that a maximum phase advance (panels D-F)
or phase delay (G-I) is induced, 
as indicated by the arrows in the phase response curves, panels A-C.

When the pulse is given at the moment that leads to a maximum phase delay,
then, in all three models,
the ADP pulse only slows down the increase of the phosphorylation level.
When the ATP level returns to 100\%, the phosphorylation level reaches the same peak height as compared
to a situation where no pulse is given.
Therefore, the subsequent dephosphorylation phase, when all KaiA is sequestered,
has the same length as when no pulse is given.
The ADP pulse thus predominantly slows down phosphorylation.

When the pulse is given at the moment that leads to a maximum phase advance,
then, in all three models, 
the ADP pulse causes the immediate start of the dephosphorylation phase.
Therefore, the peak in the phosphorylation level is lower compared to the unperturbed case,
and the number of hexamers or monomers that is able to sequester KaiA is also smaller.
The subsequent dephosphorylation phase shortens, causing the phase advance.
Only in the Rust model, the ADP pulse also initiates the immediate sequestration of KaiA, 
because, due to the sudden drop in phosphorylation rate from the S to the D state, 
KaiA is incapable anymore to prevent its own sequestration.

Clearly, the period of the Paijmans model is insensitive to changes in the average level of \ATPfrac{},
but the model is still entrainable to time-varying changes in bulk ATP levels.
If input compensation is achieved at the ensemble level,
we expect a large change in the amplitude of the phosphorylation level,
combined with large, equal but opposite changes in the lengths of the phosphorylation and dephosphorylation phases
of the oscillation, see \fref{ModelMeanATPDependence}, panels C and D.
However, even though the amplitude decreases by 40\%, the changes in the time intervals of the 
phosphorylation and dephosphorylation phases are small.
Therefore, it is not clear from the data in \fref{ModelMeanATPDependence}
how the Paijmans model achieves the high level of input compensation. 
As we show next, input compensation arises not only at the network level, 
but also the level of the individual hexamers.

\subsection*{As the ATP fraction decreases, individual hexamers go through a smaller cycle in phosphorylation state space}
Up to now, the effect of ATP on the phosphorylation level has only been studied 
at the level of the mean phosphorylation level, both experimentally and theoretically.
We want to know, for the Paijmans model, the effect of the ATP fraction 
on the phosphorylation dynamics of individual hexamers.
To this end, for each hexamer, we track in time the number of phosphorylated threonine sites, $\npT{}$,
and serine sites, $\npS{}$ , as it moves through its phosphorylation cycle (see Methods section).

In \fref{PhaseFlux_ATPcompare}, we show the distribution of hexamers in phosphorylation state space, 
\rmlabels{P}{}{\npT{},\npS{}}, and the fluxes between states,
at different phases of the oscillation.
Panel A shows, as indicated by the arrows, the time intervals at which we take data during the oscillation,
and panels B-F show the state of the ensemble during these intervals, at \ATPfrac{}=100\%.
Panels G-L show the same, but now for \ATPfrac{}=50\%.
Comparing the most populated phosphorylation states in panels D and J, 
taken at the peak of the phosphorylation level,
it is clear that more threonine and serine sites are phosphorylated at 100\% ATP level than at 50\%.
Thus, higher phosphorylation rates result in wider path through phosphorylation state space
for individual hexamers.
This is a clear signature of another mechanism of period robustness:
at higher phosphorylation rates, corresponding to higher \ATPfrac{},
the rate at which a system goes through state space increases,
yet the length of one cycle also increases \cite{Lakin-Thomas1991,Hatakeyama2015}.
It is important to note that the phenomenon of the variation of the phosphorylation cycle with the bulk ATP level 
emerges as a prediction of the Paijmans model; we did not deliberately design the model to have this property.

\subsection*{Microscopic cycles illuminate input compensation in individual hexamers}
Given our observation that individual hexamers make smaller cycles in
phosphorylation state space when \ATPfrac{} decreases, we wanted to know how this affects the
timing for switching between the active and inactive conformation.
This timing is an important factor in the period of the oscillation,
because it determines when and how long KaiA is sequestered.  To this
end, we measure the time between two important events during a full
cycle of a hexamer, illustrated in \fref{MicroscopicDynamics}A: 1) The
start of the phosphorylation cycle, when a hexamer is in the active
state and, for the first time, either a threonine site or serine site
is phosphorylated; 2) and half-way of the cycle, when the hexamer is
in the inactive state and is bound to six KaiB monomers.  For all
hexamers, we track the first passage time from the start to reaching
half-way, $\Delta\rmlabels{t}{}{active}$, and from half-way to the
start of a new cycle, $\Delta\rmlabels{t}{}{inactive}$. 
The time of a full cycle is defined as
$\Delta\rmlabels{t}{}{cycle}=\Delta\rmlabels{t}{}{active}+\Delta\rmlabels{t}{}{inactive}$.

Note that the time in the active phase, $\Delta\rmlabels{t}{}{active}$, 
not only includes the phosphorylation and the switching to the inactive state, 
but also includes the binding of KaiB monomers.
We include the binding of KaiB in this state definition,
because KaiC recrosses the dividing surface that separates the active from the inactive state many times,
before it is finally committed to the inactive state.
Including the binding of KaiB into our criterion does not affect the results,
because the rate of KaiB binding is independent of \ATPfrac{},
and any change in $\Delta\rmlabels{t}{}{active}$ due to a different \ATPfrac{} is therefore 
related to changes in phosphorylation rates.

In \fref{MicroscopicDynamics}B we show histograms of individual cycle times, $\Delta\rmlabels{t}{}{cycle}$,
at 100\% ATP (solid blue line) and 50\% (dashed orange line).
The distribution has maxima at multiples of the period of the oscillation, 
indicated by dashed vertical lines.
Clearly, the cycle times of individual hexamers coincide with the period of the oscillation.
Peaks at multiples of $\Delta\rmlabels{t}{}{cycle}$ correspond to hexamers that do not bind 6 KaiB monomers
during the first period, and hence have to wait for another round, or more,
to make the full cycle.
Since the histogram for 50\% ATP has a fatter tail, 
hexamers are more likely to wait multiple periods before completing the cycle,
showing that indeed fewer hexamers participate in an oscillation at lower ATP fractions.
\fref{MicroscopicDynamics}C shows the distributions of $\Delta\rmlabels{t}{}{active}$, 
at 100\% and 50\% ATP.
The distribution again has multiple peaks, 
mirroring those in the distribution of cycle times (panel B).
This indicates that at lower \ATPfrac{},
synchronization becomes impaired because fewer hexamers make it to the top of the cycle,
where they have 6 KaiB bound and are committed to the inactive state.

The inset of \fref{MicroscopicDynamics}C zooms in on the first peak of
the distribution, emphasizing that even though the phosphorylation
rates are different, the modes of the first passage time distribution
are remarkably similar; the difference is only 1-2 hours. 

This paradox can be resolved by noting that the switch from the active to inactive
state is determined by the difference between the number of
phosphorylated serine and threonine sites, $\npS{}-\npT{}$, respectively. 
In the Paijmans model, the phosphotransfer rates for the threonine site are much faster than for the serine site 
(see \cite{Rust2007, Nishiwaki2007, Paijmans2017}).
Therefore, compared to the slow phosphorylation of the serine sites, 
\npT{} will quickly reach its steady-state level during the phosphorylation phase. 
The steady-state level of \npT{} will thus set the number of serine sites 
that need to be phosphorylated before the hexamer can switch to the inactive state. 
The steady-state level of \npT{} decreases as the ATP fraction of the buffer is reduced, 
because the rate of phosphorylation decreases and the rate of dephosphorylation 
increases  (in the presence of KaiA) with lower \ATPfrac{}, respectively. 
Consequently, as \ATPfrac{} is decreased, less serine sites need to be phosphorylated
for the hexamer to switch conformation, which compensates for the lower rate of phosphorylation.
This reasoning implies that the levels \npS{} and \npT{}, 
at which a hexamer switches to the inactive state, also decreases. 
This can be seen in panel E and F, 
which show the distribution of phosphorylation states (\npS{},\npT{}), 
at the moment a hexamer flips from active to inactive, at
\ATPfrac{}=100\% and 50\%, respectively.  Clearly, \npS{} and \npT{}
tend to be lower at the moments of switching, when \ATPfrac{}=50\%.
This also suggests that at lower \ATPfrac{}, fewer monomers are double
phosphorylated.  \fref{MicroscopicDynamics}G shows histograms of the
phosphorylation states of the hexamers at the moment when all KaiA is
sequestered in the system for the first time during a period. 
At $\ATPfrac{}=50\%$, 40\% of the hexamers have one or more monomers in
the D state, while at \ATPfrac{}=50\%, this fraction is reduced to 10\%.

Lastly, the distribution of inactive times, 
$P(\Delta t_{\mathrm{inactive}})$ in \fref{MicroscopicDynamics}D, can be explained.
The distribution exhibits a shoulder at \ATPfrac{}=100\%, which is due to the fact that:
1) The number of hexamers that are in the inactive state is higher, 
reflected by the higher first peak in \fref{MicroscopicDynamics}B,
resulting in a longer time where all KaiA is sequestered.
Therefore, the time hexamers have to wait before another round of phosphorylation starts increases,
which is included in $\Delta t_{\mathrm{inactive}}$;
2) Hexamers start their inactive phase at a higher phosphorylation level,
comparing panels E and F in \fref{MicroscopicDynamics},
which results in a longer dephosphorylation phase.

We found that in the Paijmans model, input compensation is present both at the ensemble level 
as well as at the level of an individual hexamer.
To confirm that the antagonistic effect of phosphorylation of the threonine and serine site on the conformation
indeed gives input compensation at the level of a single hexamer,
we will now study a simpler model which can only achieve input compensation through this mechanism. 
As we will show, this minimal model not only describes input compensation, but also entrainability.
We emphasize that while this model can yield phase-response curves that are qualitatively similar 
to those observed experimentally, 
the main purpose of the simplified model is to clarify the physical origins of input compensation and entrainability 
at the level of a single hexamer, not to quantitatively fit measurements of the Kai system.

\subsection*{Simple model with antagonism shows input compensation and entrainability at hexamer level}
To illuminate the mechanism behind input compensation at the hexamer level, 
we here introduce a simplified version of the Kai system, 
consisting of only one hexamer, flipping between the active and inactive conformation, without KaiA and KaiB.
This way, we can study hexamer level input compensation,
without input compensation at the ensemble level resulting from the interaction between hexamers via KaiA.
In this simple model, shown in \fref{SimpleModel}A, each monomer has a threonine and serine site, 
and the phosphorylation of these sites have an equal but opposite effect on the free energy difference 
between the two conformations: $\Delta\rmlabels{G}{conf}{I,A}=\delta\rmlabels{g}{conf}{}(\npS-\npT)$.
The resulting flipping rates between the active and inactive state are
$\rmlabels{k}{conf}{f/b}=\rmlabels{k}{conf}{0}\,{\rm exp}(\mp0.5\Delta\rmlabels{G}{conf}{I,A})$,
where \rmlabels{k}{conf}{0} is a prefactor setting the overall flipping rate.

Phosphorylation and dephosphorylation occur through phosphotransfer with ATP and ADP in the nucleotide
binding pocket, respectively.
In the active state, the nucleotide exchange and hydrolysis rates are much higher than the phosphotransfer rates,
such that the occupancy of the nucleotide binding pocket with
ATP compared to ADP is given by the steady state fraction \rmlabels{\beta}{}{ATP}.
We include hydrolysis of ATP in the binding pocket because ADP is necessary for dephosphorylation.
In the inactive state, the nucleotide exchange rate is zero, while hydrolysis rate is fast, 
such that the binding pocket is always occupied with ADP. 
This means, like for the Kai system, dephosphorylation in the inactive state is independent of \ATPfrac. 
We find \rmlabels{\beta}{}{ATP} in the active state by solving the steady state equation for
nucleotide exchange between the bulk and the binding pocket, 
where we assume that the association rates of ATP and ADP are equal and much higher than the dissociation rates \cite{Paijmans2017},
\begin{equation}
\rmlabels{\beta}{}{ATP} =
\begin{cases}
\frac{\ATPfrac}{\ATPfrac + \rmlabels{K}{}{ATP/ADP}(1-\ATPfrac) + \gamma} & \text{active,}
\\
0 & \text{inactive.}
\end{cases}
\end{equation}
Here, \rmlabels{K}{}{ATP/ADP} is the dissociation rate constant of ATP divided by that of ADP and $\gamma$
the hydrolysis rate constant divided by the ADP dissociation rate. 
The phosphotransfer rates between the nucleotide and one of the phosphorylation sites are given by
$\rmlabels{k}{}{XY}=\rmlabels{\beta}{}{ATP}\,\rmlabels{k}{0}{XY}$ and 
$\rmlabels{k}{}{YX}=(1-\rmlabels{\beta}{}{ATP})\,\rmlabels{k}{0}{YX}$,
where X and Y are one of the four possible phosphorylation states of a monomer U,T,D and S.
Note that both in this simple model and in the full model \cite{Paijmans2017}, 
the dephosphorylation rate depends on the presence of ADP in the binding pocket,
which in turn makes it dependent on the bulk ATP fraction.
This dependence will affect the steady-state level of the number 
of phosphorylated threonine sites at a given \ATPfrac{},
and the cycle the hexamer traverses in phosphorylation state-space.
However, this change of the dephosphorylation rate is not essential for input compensation through the 
adaptation of the phosphorylation cycle.
In fact, the same model behavior could have been achieved 
through an appropriate dependence of only the phosphorylation rate on \ATPfrac{}.
Indeed, the absolute rates do not matter: only how the relative rates of phosphorylation and dephosphorylation depend 
on the ATP fraction is important for the behavior of both the full model and the minimal model presented here.

To elucidate the function of the antagonism in the antagonism model, 
we compare it with another minimal model, the fixed cycle model as shown in \fref{SimpleModel}B, 
where each monomer only has one phosphorylation (serine) site.
In this model, the hexamer flips to the inactive state when $\npS\geq5$ 
and back to the active state when $\npS\leq1$, with a rate \rmlabels{k}{conf}{0}. 
The phosphotransfer rates and their dependence on \ATPfrac{} are the same as for the antagonism model.
Parameters of the antagonism and fixed cycle models are given in \tref{SimpleModelParameters}.

\begin{table}[th]
 \begin{tabular}{l r r| l r r}
  {\bf Parameter}          & {\bf Value} & & {\bf Parameter}    & {\bf Value} & \\ \hline \hline
  \rmlabels{k}{}{UT},\rmlabels{k}{}{SD} & 20.0 & \rmlabels{h}{-1}{} & \rmlabels{K}{}{ATP/ADP}      &  0.1 & \\
  \rmlabels{k}{}{TU},\rmlabels{k}{}{DS} & 10.0 & \rmlabels{h}{-1}{} & \rmlabels{k}{conf}{0}        &  0.2 & \rmlabels{h}{-1}{} \\
  \rmlabels{k}{}{TD},\rmlabels{k}{}{US} &  0.4 & \rmlabels{h}{-1}{} & $\delta\rmlabels{g}{conf}{}$ &  2.0 & kT \\
  \rmlabels{k}{}{DT},\rmlabels{k}{}{SU} &  0.2 & \rmlabels{h}{-1}{} & $\gamma $	             &  0.25 & \\                              
 \end{tabular}
 \caption{\tlabel{SimpleModelParameters} Parameters used in the simple antagonism model and the fixed cycle model.}
\end{table}

We define the time the hexamer needs to progress through a cycle, $\Delta\rmlabels{t}{}{cycle}$,
as the interval between two consecutive conformational switches to the active state. 
Just like for the full model, measuring cycle times is hampered
due to the fact that the hexamer recrosses the dividing surface between the two conformations many times
before finally committing to a conformation.
However, instead of using a grace interval to integrate out the recrossings \cite{Becker2012}, 
it turns out that choosing a low rate for the conformational switch, \rmlabels{k}{conf}{0}, 
compared to the phosphorylation rates is already sufficient to prevent recrossing.

\fref{SimpleModel}, panels C and D, show histograms of $\Delta\rmlabels{t}{}{cycle}$ 
for the antagonism and fixed cycle model, respectively.
Both panels show histograms at a 100 and 50\% ATP buffer.
Remarkably, for the antagonism model, when \ATPfrac{} is lowered to 50\%, 
the mode of the histogram shifts to a lower cycle time.
This behavior is surprising;
We would expect the cycle times to increase since all the phosphorylation rates 
are lowered by a factor \rmlabels{\beta}{}{ATP}.
For the fixed cycle model, the cycle times do indeed increase when \ATPfrac{} is lowered.
In \fref{SimpleModel}E we show how the mode of $P(\Delta\rmlabels{t}{}{cycle})$ 
changes with \ATPfrac.
Clearly, changing the ATP fraction has a stronger and opposite effect on the cycle time of the fixed cycle model
as compared to the antagonism model. 

To highlight how the phosphorylation behavior changes with \ATPfrac{} in the antagonism model,
we show a histogram of the phosphorylation states where the hexamer changes conformation, 
$P(\npT,\npS)$, \fref{SimpleModel}F, for different bulk ATP levels.
At any given ATP level the distribution shows two peaks: 
One in the lower left corner resulting from the conformational switch to the active state which does not change with \ATPfrac,
and one for the conformational switch to the inactive state which does depend strongly on \ATPfrac.
As we lower \ATPfrac, starting top right, the mode of $P(\npT,\npS)$ for switching to the inactive state 
gradually moves along the diagonal to lower phosphorylation levels.
This shows that the antagonism model is able to adapt its cycle through state space 
to the external ATP fraction.
Just like in the original Kai model, the antagonism model compensates the lower phosphorylation rates with a smaller 
cycle through state space.
For the fixed cycle model, the phosphorylation state at which the hexamer changes conformation
does not change with \ATPfrac{}, such that the cycle through state-space is fixed.
Therefore, as the phosphorylation rates decrease, the time required for a full cycle can only increase in this model.

How does the hexamer in the antagonism model adapt its phosphorylation dynamics to the bulk ATP fraction?
\fref{SimpleModel}G shows a cartoon of the paths a hexamer takes through phosphorylation state-space 
at 100\% and 50\% bulk ATP fractions.
Starting in the lower left corner where the active hexamer starts unphosphorylated,
the phosphorylation level of the threonine sites will quickly reach their steady-state level, $\bar{\mathrm{n}}_{\mathrm{T}}$,
because phosphotransfer with the threonine sites is much faster than with the serine sites.
This steady-state level increases with \ATPfrac, 
such that at 100\% ATP an active hexamer will typically have more threonine sites phosphorylated than at 50\% ATP.
Because a hexamer switches to the inactive state when $\npS>\npT$,
the number of serine sites that need to be phosphorylated before the hexamer can switch
to the inactive state depends on $\bar{\mathrm{n}}_{\mathrm{T}}$, which in turn is set by \ATPfrac.
As the phosphorylation of a serine site is slow, 
the longer trajectory at $\ATPfrac=100\%$ will increase the number of slow steps required 
to switch to the inactive state, and the number of slow steps to switch back to the active state again.
This negates the effect of the faster phosphorylation rates.

Next to input compensation for a constant bulk ATP fraction,
the antagonism model also shows entrainability at the level of a single hexamer by a transient change in the ATP fraction.
The blue arrow in panel G shows how a phase advance can occur when temporarily \ATPfrac{} is lowered
from 100\% to 50\% ATP.
Because the phosphorylation level of the threonine sites will quickly adjust to the lower steady-state level 
for the new \ATPfrac,
the number of phosphorylated serine sites will suddenly outnumber the threonine sites,
allowing the hexamer to change conformation earlier, leading to a phase advance.
In contrast, when the transient pulse of a lower ATP level is given near the beginning of the cycle,
the conformational switch to the inactive conformation is delayed.
In this case, lowering \ATPfrac{} will slow down the phosphorylation of the serine sites, 
slowing down the progress in the \npS{} direction. 
Lowering \ATPfrac{} will lower $\bar{\mathrm{n}}_{\mathrm{T}}$, 
but because near the beginning of the phosphorylation cycle \npS{} is low, 
the system will not immediately switch to the inactive state. 
When \ATPfrac{} returns back to 100\% at the end of the pulse 
and the hexamer has not switched to the inactive state yet,
the hexamer will return back to its old trajectory, 
i.e. the same trajectory when no pulse had been given. 
However, the pulse of lower \ATPfrac{} means that it will have progressed less far in the \npS{} direction, 
and this leads to a phase delay. 
A pulse of lower \ATPfrac{} will thus lead to a phase advance when the pulse induces a premature switch, 
which is what tends to happen when the pulse is given later in the phosphorylation phase of the cycle, 
while it will lead to a phase delay when the pulse does not induce a switch, 
which is the typical scenario when the pulse is given earlier in the cycle.
This simple model thus qualitatively explains how input compensation and entrainability 
can be achieved simultaneously through an antagonism mechanism within individual hexamers 
and thus shows that input compensation at a network-wide level is not essential.

\section*{Discussion}
All circadian clocks have to fulfill two
seemingly conflicting requirements in order to be a good predictor of time: 
A robust circadian period under a wide range of external conditions, 
and entrainability such that the clock always moves in phase with the day-night cycle.  
It is important to note that most circadian clocks found in nature
have a free running period that is one to a few hours off from 24 hours.
The idea is that this mismatch improves the entrainability of the clock, 
which is necessary to keep a robust phase relation with the day 
in the presence of inevitable biochemical noise \cite{Pfeuty2011, Monti2017}.
Therefore, when comparing different models of circadian clocks,
it is crucial to test them on the basis of both input compensation \emph{and} entrainability.
It is becoming clear that the daily shifts in the metabolic state of organisms, ranging from cyanobacteria to mammals,
are an important Zeitgeber for their circadian clocks \cite{Stokkan2001,Rust2011,Woller2016}.
Recent experiments show that the daily shifts in the cytoplasmic ATP level of the cyanobacterium \textit{Synechococcus elongatus}
are an important cue for the entrainment of its Kai circadian clock \cite{Pattanayak2015}.
To find out how changes in the ATP level affect the Kai oscillator,
we compared two widely-used models of the post\hyp{}translational
oscillator, the hexamer model by Van Zon \textit{et. al.} \cite{VanZon2007} and the
monomer model by Rust \textit{et. al.} \cite{Phong2012}, with the Paijmans model \cite{Paijmans2017}, 
and studied how well they fulfill the robustness and entrainability criteria.

We find, in agreement with experiments \cite{Phong2012}, 
that the period in the Paijmans model is almost unaffected by the
bulk ATP fraction, and that its hard to determine from the mean
quantities related to the phosphorylation level what sets this period stability. 
Apart from the amplitude in the phosphorylation level,
other quantities such as the length of the phosphorylation and
dephosphorylation phase, do not change much with the bulk ATP fraction
in the Paijmans model.  The other two models show clear signatures of input
compensation.  The Van Zon model has too little input compensation,
however, as shortening the dephosphorylation phase by sequestering
less KaiA is unable to compensate slower phosphorylation.  On the
other hand, the Rust model has too much input compensation, since the
period decreases with lower ATP. Lowering the phosphorylation rates
has a too strong effect on the positive feedback loop regulating KaiA
sequestration, shortening the period too much. 
Note however, that we chose the affinity for ATP over ADP a factor 10 higher 
to get good agreement with experiments for the sensitivity of the phosphorylation speed to 
the bulk ATP fraction \cite{Paijmans2017}.
The fact that the Paijmans model is so stable, is therefore perhaps not so surprising, 
as the sensitivity to ADP in the bulk is a factor 10 lower in the Paijmans model compared to the
other two.

We then checked the entrainability of the respective models, and found
that the entrainability of the Paijmans model is comparable to that of the
other models, even though the relative affinity of ADP versus ATP is
lower in the Paijmans model.  When we make the relative ADP/ATP sensitivity in
the Van Zon and Rust models similar to that of the Paijmans model, the
amplitudes of the phase response curves in these other models become
very small, leading to poor entrainability.  This showed that the Paijmans
model, contrary to previous models, is capable of maintaining a robust
circadian period, while at the same time being strongly entrainable.

To elucidate how the Paijmans model achieves this combination of period
robustness and good entrainability, we studied the phosphorylation
cycle of the threonine and serine sites in individual hexamers, at
100\% and 50\% bulk ATP fractions.  This showed that at 50\% ATP, when
effective phosphorylation rates are lower, individual hexamers go
through a smaller phosphorylation cycle.  Our analysis also revealed
that the distribution of times for hexamers to complete a full cycle
peaks at multiples of the period and that peaks at times higher than
the period become more pronounced at lower ATP fraction, as fewer
hexamers make it through the full cycle each period.  Remarkably, the
time required to complete the first part of the cycle, between when
phosphorylation starts and when 6 KaiB monomers are bound, seems to be
little affected by the bulk ATP fraction.  This leads to the question
of how, at different effective phosphorylation rates, the timing of
the conformational switch that ends the first part of the cycle can be almost unaffected.

To address this question, we made histograms of the phosphorylation
states of the hexamers when they switch to the inactive state.  At
lower phosphorylation rates, hexamers switch to the inactive state at
a lower phosphorylation level, compensating the longer time they
need to phosphorylate the sites. 
This is in marked contrast with
previous models by Van Zon \textit{et. al.} and Rust \textit{et. al.},
where the time required for individual units (be it hexamers or monomers) 
to complete a full cycle, can only increase with decreasing ATP fractions.  
In these models, period stability can only be achieved
through the interaction {\em between} the individual hexamers via KaiA
sequestration \cite{Hatakeyama2015}, 
that is, via the delay between the moment where KaiC no
longer needs KaiA to progress along the
cycle and the point where KaiC sequesters KaiA: 
At lower \ATPfrac{}, less KaiC makes it to the top of the cycle during the delay, 
making the dephosphorylation phase shorter, which can then counteract the
longer phosphorylation phase \cite{Phong2012}.

The Paijmans model presented here exploits this mechanism too, yet also
employs another one, which acts at the level of the individual hexamers. 
This mechanism utilizes the ordered phosphorylation of the threonine and serine site in the monomers, 
in combination with their antagonistic effect on the conformation of the hexamer.  
Previously, Lin {\it et. al.} argued that this antagonism creates an ultra-sensitive switch
which provides robustness against a varying KaiA concentration \cite{Lin2014}.
However, this antagonism also plays a key role in our mechanism of input compensation.
The antagonism entails that the state at which the hexamer switches conformation 
does not depend on the absolute number of phosphorylated threonine and serine residues, 
but rather on their relative amounts. 
When the ATP fraction is low, their are typically fewer serine and threonine sites phosphorylated 
(see \cite{Rust2011, Paijmans2017}),
but since the switch to the inactive conformation does not depend on the 
absolute number of phosphorylated sites,
but on their relative amounts,
hexamers switch conformation at a lower number of phosphorylated sites.
The net result is that, although the cycle in state space is smaller at lower ATP fraction,
the moment a hexamer switches is robust against changes in the ATP fraction.
And, since the phosphorylation rates \emph{are} sensitive to changes in the fraction,
the hexamer still undergoes a phase shift, essential for entrainability.
We thus propose a new function for the ordered phosphorylation cycle of the KaiC monomers: 
It allows the oscillator to combine period robustness with high entrainability.
We end our study with a simplified model of the hexamer 
which includes the antagonism but no interaction between the hexamers through KaiA.
This simple model suggests that the antagonism is enough to achieve input compensation  
and entrainability and that a network-wide mechanism is not essential.

This mechanism of input compensation is a result which we identified after carefully studying the 
phosphorylation dynamics of individual hexamers and is not an assumption built into the Paijmans model.
Indeed, the mechanism was an outcome of the model, 
which we had developed with the aim to describe the wealth of data that is available for this system \cite{Paijmans2017}. 
The antagonism extends our understanding of the robustness of the 
post-translational Kai oscillator beyond its implications for the robustness against varying nucleotide concentrations.
Apart from making the oscillator more robust against extrinsic fluctuations such as in temperature and protein concentrations \cite{Lin2014},
the mechanism will also make the oscillator more robust against intrinsic fluctuations arising from stochasticity in biochemical reactions.
Due to this intrinsic noise, the hexamers in the ensemble will generally have different phosphorylation levels.
Some hexamers, the front runners, will take a relatively large cycle.
When these hexamers reach the dephosphorylation phase, they will sequester KaiA, taking it away from the laggards 
that are still in the phosphorylation phase of the cycle. 
In the Van Zon model, this stops the progression of the laggards around the cycle, 
causing them to fall back to the beginning of the cycle. 
In contrast, in the Paijmans model discussed here, the laggards will make a shorter cycle, 
but they will nonetheless complete it and hence subsequently participate in sequestering KaiA. 
This increases the number of hexamers that participate in the oscillations, 
and more specifically, in sequestering KaiA. 
This is likely to increase the robustness.

Recently, it was shown that the transcription factor Elk-1 is first activated via the fast phosphorylation 
of certain residues by the kinase ERK, 
and later deactivated via the slower phosphorylation of other residues by ERK \cite{Mylona2016}. 
Phosphorylation at different rates thus facilitates the robust transient activation of Elk-1 in response to ERK activation.
This suggests that our results on cyclic phosphorylation of a protein combined with an antagonistic effect 
of phosphorylation on its activity, may have consequences even outside the realm of circadian clocks.

\subsection*{Experimental verification}
Here we explore the possibilities for experimentally verifying the prediction that individual hexamers
adapt their phosphorylation cycle to the ATP fraction in the bulk.
In the original publication \cite{Paijmans2017}, we identify two key predictions of our model
and we describe experiments to verify them:
1) The relative stability of the two conformations is determined by the ATP fraction in the binding pockets of the 
CI domain and 2) this fraction is set by the relative number of phosphorylated serine and threonine sites 
in the CII domain of the hexamer. 
The verification of these predictions would already suggest input compensation at the hexamer level since 
both conditions are necessary for the mechanism.

However, to directly measure the adaptation of the phosphorylation cycle, 
we would have to measure the phosphorylation state of all hexamers at different phases of the oscillation,
at different ATP levels (see \fref{PhaseFlux_ATPcompare}).
Current experimental techniques, such as mass spectrometry, SDS-PAGE \cite{Rust2007,Nishiwaki2007}, and time resolved fluorescence spectroscopy \cite{Murayama2011},
can identify the four (U,T,D and S) phosphorylation states of the KaiC monomers. 
Experiments show, in agreement with our model, that as the bulk ATP level decreases, the amplitude of the phosphorylation cycle decreases \cite{Rust2011, Phong2012}.
However, the measured decrease in the amplitude can result from two scenarios:
1) Hexamers go through a fixed phosphorylation cycle such that at a lower bulk ATP level, 
when the amplitude in the mean phosphorylation level decreases, fewer hexamers traverse a full cycle. 
Therefore, fewer hexamers bind KaiB and sequester KaiA during an oscillation.
2) Hexamers traverse a smaller phosphorylation cycle at a lower ATP level, such that the fraction of hexamers 
that traverse a full cycle during an oscillation decreases much less than the decrease in the amplitude
of the mean phosphorylation level.
Current experiments do not allow us to distinguish between these two scenarios.
In the future, chemical crosslinking of KaiC hexamers and the indentification of crosslinked 
complexes by mass spectrometry may allow us to track the 
phosphorylation state of whole hexamers \cite{Leitner2016,Snijder2017}.
This would make it possible to track the fraction of hexamers in a certain phosphorylation state as the 
Kai system traverses its phosphorylation cycle.

Less direct tests of our conclusions, in contrast, might be possible with current technology.  
In particular, our model predicts that the switch to the inactive confirmation depends on the difference between
the number of phosphorylated threonine and serine sites and not on their absolute level.
This allows hexamers to traverse a smaller phosphorylation cycle and still switch to the inactive conformation, 
bind KaiB and sequester KaiA.
Therefore, by verifying that hexamers switch to the inactive state at different phosphorylation levels, 
we indirectly test whether hexamers go through a smaller cycle.
We propose two experiments, based on currently available techniques,
that can test this prediction:

1) Mix different ratios of (phosphomimetic) monomers in the T and S states,
and measure the fraction of hexamers that are in the inactive state, as was done in \cite{Kitayama2013,Lin2014}.
Now repeat this experiment at different ratios of unphosphorylated monomers to phosphorylated monomers (at a specific ratio of T vs S).
The prediction is that the key variable that determines the fraction of hexamers that are in the inactive conformation
is $\npS{}-\npT{}$, and not the ratio of unphosphorylated to phosphorylated monomers.
To find the fraction of hexamers in the inactive state, one could measure it directly \cite{Oyama2016},
or indirectly via the fraction of KaiC bound to KaiB \cite{Phong2012, Lin2014}.

2) The same setup as in the experiment above, but now retrieve those monomers from solution 
that are bound to KaiB using coimmunoprecipitation, as was done in \cite{Lin2014}.
Then measure the distribution of phosphorylation states of these monomers.
If KaiC were to bind KaiB only in specific phosphorylation states 
(e.g., hexamers only bind KaiB when all monomers are fully phosphorylated), 
contrary to our prediction, then the distribution of phosphorylation states of the monomers bound to KaiB 
would not change substantially with different initial ratios of phosphorylated to unphosphorylated KaiC.
However, when the ability to bind to KaiB depends on the difference between S and T state monomers in a hexamer, 
as our model predicts, then, we expect the fraction of unphosphorylated monomers to scale 
with the initial fraction of unphosphorylated (U-state) monomers.

\section*{Methods}

\subsection*{Models used}
We use the Rust model described the SI of \cite{Phong2012} and the Van Zon model in the SI of \cite{VanZon2007},
both described with ordinary differential equations (ODE's), 
propagated using the NDSolve function of Mathematica 8 (Wolfram Research).
The Paijmans model, introduced in \cite{Paijmans2017}, 
is propagated using the dedicated Monte Carlo algorithm described in the same paper.

\subsection*{Model sensitivity to bulk ATP fraction}
We want to compare the Van Zon and Rust models, using a relative affinity for ATP versus ADP,
such that the change of the ATP fraction in the nucleotide binding pockets given that KaiA is bound, 
due to changes in \ATPfrac{}, is similar to the change in the Paijmans model.
In the Paijmans model, the steady state fraction of ATP in the CII binding pocket, \rmlabels{\beta}{CII}{ATP},
given that KaiA is always bound to CII, is given by
\begin{equation}
 \elabel{betaCIIATP}
 \rmlabels{\beta}{CII}{\mathrm{ATP}} = \frac{\ATPfrac{}\,\rmlabels{k}{CII\cdot ADP}{off, KaiA}}{ \ATPfrac{}\,\rmlabels{k}{CII\cdot ADP}{off, KaiA} + \rmlabels{k}{CII}{hyd} + (1-\ATPfrac{})\,\rmlabels{k}{CII\cdot ADP}{off, KaiA}\rmlabels{K}{CII}{ATP/ADP}}.
\end{equation}
Here, \rmlabels{k}{CII\cdot ADP}{off, KaiA} and \rmlabels{k}{CII}{hyd} are the dissociation rate of ADP when KaiA is bound 
and the hydrolysis rate of ATP, in the CII domain of KaiC, respectively.
Using the parameters presented in \cite{Paijmans2017},
we find that $\rmlabels{\beta}{CII}{\mathrm{ATP}}$ decreases with 15\% as we lower \ATPfrac{} from 100\% to 50\% in \eref{betaCIIATP}.
The ATP fraction in $\beta_{\mathrm{ATP}}$, Eqn. 1,
has a similar scaling when we set $\rmlabels{K}{}{ATP/ADP}=0.19$.

\subsection*{Calculating mean quantities in Fig. 2}
Results shown in Fig. 2 were taken after 10 oscillations, 
such that the system has reached steady state oscillations.
For the Paijmans model, presented quantities are averages over 400 consecutive oscillations.
The period is defined as the mean of the peak-to-peak time in the phosphorylation level.

To calculate the flux trough a cycle, Fig. 2E, 
we require a reaction that has to take place at least once,
in order for a hexamer or monomer to complete a full cycle.
Here, a full cycle is defined as a series of states the hexamer or monomer has to go through 
in order to be able to sequester KaiA.
For the Rust model, we calculate the flux between the U and T phosphorylation states, 
integrated over a period $P$:
\begin{equation}
 Q_{\mathrm{Rust}}=\frac{1}{P} \int_0^{P}\left(k_{\mathrm{UT}}(t)\,U(t)-k_{\mathrm{TU}}(t)\,T(t)\right)dt,
\end{equation}
where $k_{\mathrm{UT}}$ and $k_{\mathrm{TU}}$ are the time dependent (depend on the free KaiA concentration) 
rates for phosphorylation and dephosphorylation of the T state,
and $U(t)$ and $T(t)$ are the concentrations of U and T phosphorylated monomers.
For the Van Zon model, we calculate the flux between switching to the inactive state
\begin{equation}
 Q_{\mathrm{VanZon}}=\frac{1}{P} \int_0^{P}\left( k_{\mathrm{fw}}\,\mathrm{C}_6(t)-k_{\mathrm{bw}}\,\mathrm{\tilde{C}_6}(t)\right) dt,
\end{equation}
where $k_{\mathrm{fw}}$ and $k_{\mathrm{bw}}$ are the rate constants for switching to the inactive or active state,
respectively.
In the Paijmans model we measure the flux by counting the number of hexamers that 
have six KaiB monomers bound at some point during the period, averaged over 400 oscillations.

The time interval in which all KaiA is sequestered, Fig. 2F, 
is defined as the time interval when more than 99\% of all KaiA dimers are sequestered by KaiC.

\subsection*{Calculating phase response curves in Fig. 3}
To generate the phase response curves (PRC) shown in Fig. 3A-C,
we applied a 6 hour step-wise decrease in the ATP fraction.
For the ODE models, 
we derived the induced phase change by comparing the time of second through in the phosphorylation level after the onset of the pulse, 
with the same trough of the control where no pulse is given.
For the Paijmans model, which contains stochasticity,
we fit a sinusoidal function to three oscillations in the time trace of the phosphorylation level,
one oscillation after the pulse is given.
We compare this with a fit to the phosphorylation level in a control simulation, where no pulse is given,
and calculate the phase shift between the fits.
To suppress the intrinsic number fluctuations present in the Monte Carlo simulations, 
we used a simulation volume of 6 cubic microns, which is three times as large as the original volume.
Finally, to get the PRC in panel A, we averaged over 10 different trajectories for each pulse start.
The shaded region in Fig. 3A indicates the standard deviation in the phase shift of these 10 runs.

\subsection*{Microscopic dynamics of the hexamer, Fig. 4, Fig. 5 and Fig. 6}
To get the snapshots of the distribution of hexamers in phosphorylation state space shown in Fig. 4,
we first need to register the time of the troughs in the phosphorylation level, 
\rmlabels{t}{tr}{i}, for a time trace containing over 400 oscillations.
This allows us to calculate the trough-to-trough time, 
$\rmlabels{T}{tr}{i}=\rmlabels{t}{tr}{i+1}-\rmlabels{t}{tr}{i}$, for each oscillation $i$. 
Then we define time intervals of one hour for the whole trajectory, 
$(\rmlabels{t}{tr}{i} + \gamma\,\rmlabels{T}{tr}{i}, \rmlabels{t}{tr}{i} + \gamma\,\rmlabels{T}{tr}{i} + 1.0)$,
where $\gamma\in[0,1)$ sets the phase of the period at which the time intervals start.
During these intervals we keep track of the number of phosphorylated threonine sites, \npT{}, 
and serine sites, \npS{}{}, in each hexamer.
This allows us to calculate the occupancy of each phosphorylation state, $P_{\npT{},\npS{}}$,
and the fluxes between these states, during the one hour time window.

\section*{Acknowledgments}
  This work is part of the research programme of the Netherlands Organisation for Scientific Research (NWO).
  (JP and PRtW), and by NSF Grant DMR-1056456 (DKL).
  
\begin{figure}[t!]
{\centering
\includegraphics[scale=1.0]{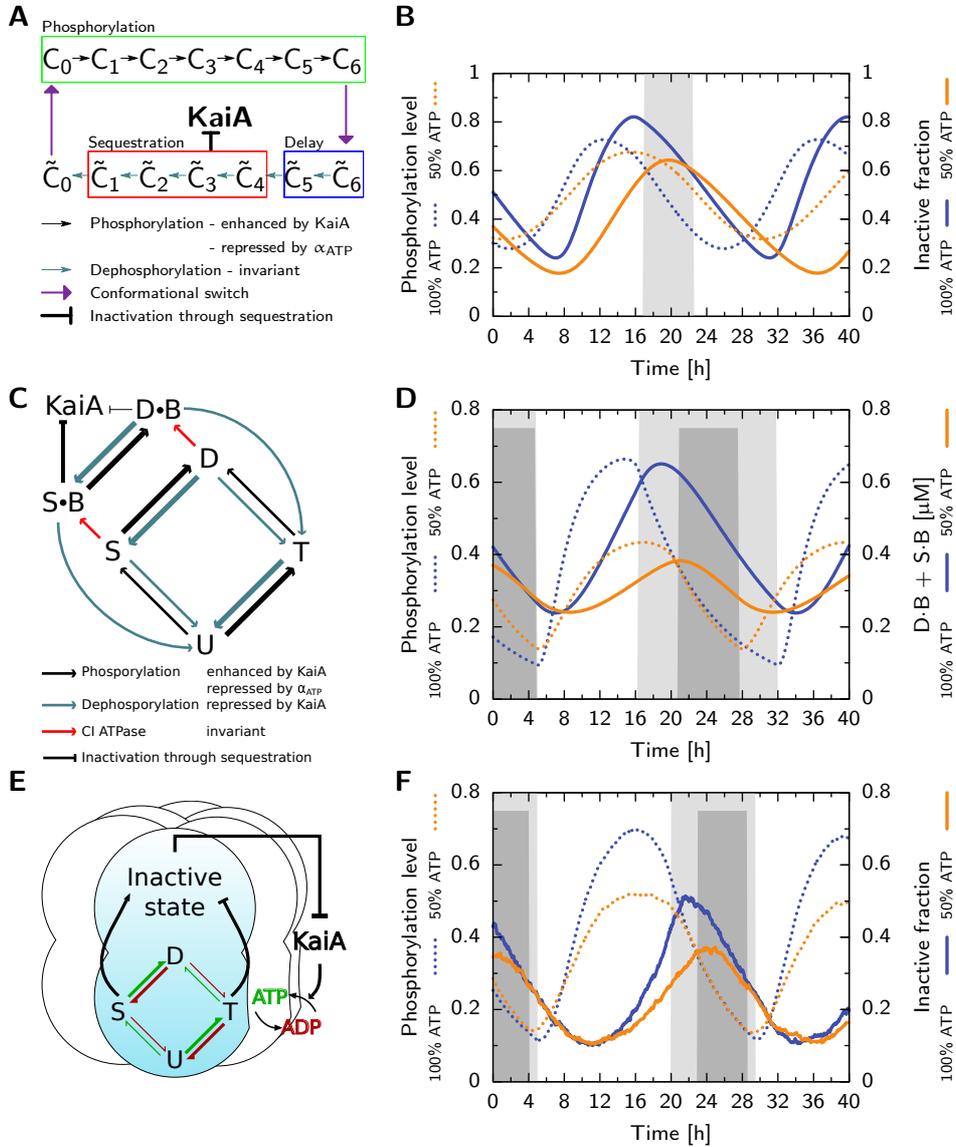}
}
\caption{\flabel{StabilityMechanisms} The different models of the Kai
  system employ different mechanisms of period stabilization. (A,B):
  Van Zon model; (C,D): Rust model (panel C after \cite{Phong2012}); (E,F): new model. 
  Phosphorylation level (dotted lines,B,D,F) and fraction of KaiC in inactive state
  (solid lines,B,F) or in S$\cdot$B + D$\cdot$B (D, solid lines) at \ATPfrac=100\%
  (blue) and  50\% (orange). Shaded regions indicate the phase where all KaiA is
  sequestered, at \ATPfrac{}=100\% (gray) and 50\% (dark gray).  
}
\flushright \emph{(continued next page)}
\hrule 
\end{figure}
\begin{figure}[t!]
\ContinuedFloat
\caption{(\emph{continuing from previous page}) 
 (A) Van Zon model: Hexamers phosphorylate in the active phase (green box),
 with a rate that depends on \ATPfrac{} and the amount
 of free KaiA, switch to the inactive state where they dephosphorylate 
 and, after a delay (blue box), sequester KaiA (red box).
 (B) Both the phosphorylation level and the fraction of
 inactive hexamers rise slower at \ATPfrac{}=50\%,
 but due to the delay between phosphorylation and KaiA sequestration,
 the dephosphorylation phase starts at a lower maximum of the inactive fraction, 
 which shortens the time window where all KaiA is sequestered.
 (C) Rust model: Monomer is predominantly phosphorylated in the order U,T,D,S. 
 Phosphorylation depends on the ATP fraction and KaiA (black arrows), 
 dephosphorylation only on KaiA (blue arrows). 
 To emphasize that phosphorylation of the threonine site is much faster than for the serine site,
 the arrows indicating (de)phosphorylation of the threonine site are thicker.
 The delay is set by the slow KaiB binding step (red arrows), 
 and mutual inhibition between KaiA and the sequestration state S$\cdot$B.
 (D) Because at \ATPfrac{}=50\% the rate from S to D is smaller,
 the ensemble starts the dephosphorylation phase at a lower concentration of KaiB-bound KaiC.
 This shortens the dephosphorylation phase, which compensates for slower phosphorylation.
 (E) New model: In each hexamer, monomers go through the ordered phosphorylation cycle,
 where T stabilizes the active state, and S the inactive state.
 Phosphorylation can only occur with ATP (green arrows) and dephosphorylation only with ADP (red arrows)
 in the CII binding pocket.
 (F) The inactive fraction (solid lines) increases slower at  \ATPfrac{}=50\%, 
 but reaches a lower maximum, leading to a shorter dephosphorylation phase.
 However, the effect on the duration of the KaiA sequestration phase 
 is not as strong as in the Rust model.
 Because in the new model the switch to the inactive state depends
 not on the absolute number of serine and threonine phosphorylated residues, 
 but rather on their difference, 
 the decrease in the phosphorylation level (dotted lines) has a smaller 
 impact on the inactive fraction (solid lines), as compared to the impact on the
 S$\cdot$B+D$\cdot$B fraction in the Rust model (panel D). 
 (Our simulations of the Van Zon and Rust models employ ordinary differential equations,
 but the new model can only be described by stochastic simulations \cite{Paijmans2017}.
 Hence the curves in F show a small amount of noise that is absent in B and D.)
}
\smallskip
\hrule 
\end{figure}

\begin{figure}[t!]
{\centering
\includegraphics[scale=1.0]{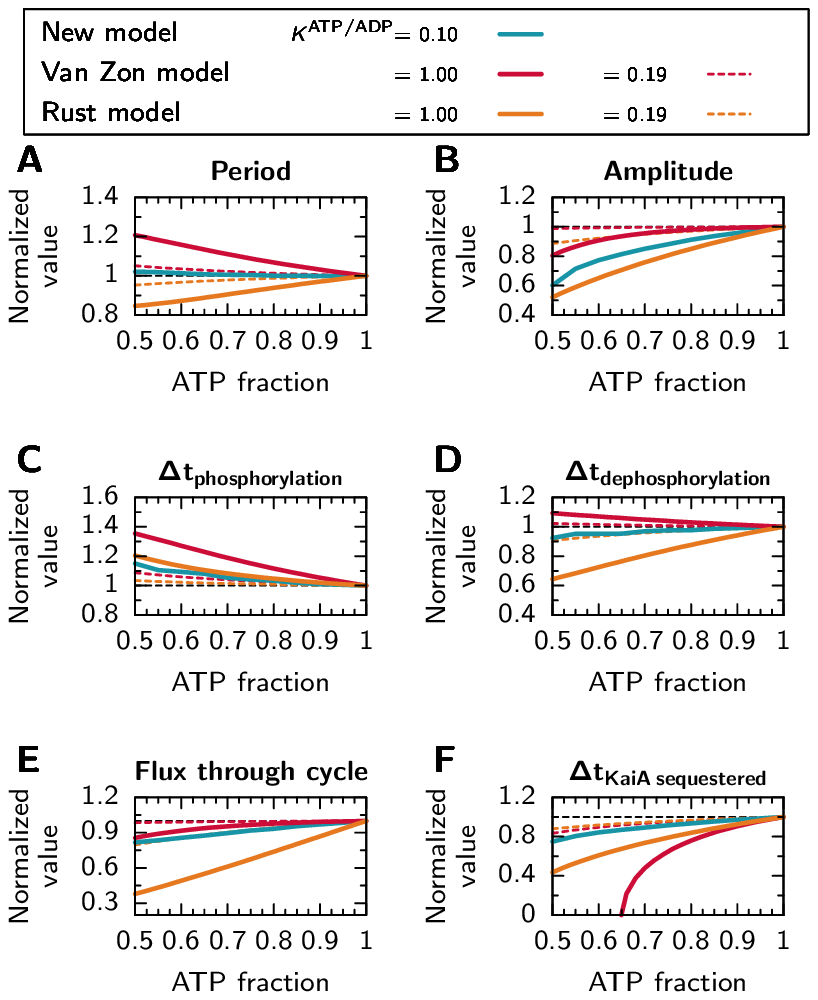}
}
\caption{\flabel{ModelMeanATPDependence} 
Dependence of the Paijmans (blue solid lines), Van Zon (red) and the Rust (orange) models on the bulk ATP fraction, 
\ATPfrac{}. The quantity being plotted is indicated by the legend above each panel,
and in each case values are normalized by the value at $\ATPfrac=100\%$,
indicated by the horizontal dashed line. 
For the Van Zon and the Rust models, we study versions with equal sensitivity for ATP and ADP, 
\rmlabels{K}{}{ATP/ADP}=1.0 (solid lines), and a smaller sensitivity for ADP, \rmlabels{K}{}{ATP/ADP}=0.19 (dashed lines).
(A) The ATP fraction has an opposite effect on the period in the Van Zon and the Paijmans model 
compared to the Rust model,
(B) but the amplitude increases with increasing \ATPfrac{} in all models.
(C) The time between a through and the next peak in the phosphorylation level, $\Delta\rmlabels{t}{}{phosphorylation}$,
decreases with \ATPfrac{} in all models, because the phosphorylation rates increase.
(D) In both the Paijmans model and the Rust model, the time between a peak and the next trough, 
$\Delta\rmlabels{t}{}{dephosphorylation}$, 
increases with \ATPfrac{}, which compensates for the faster phosphorylation rates.
In the Van Zon model however, the dephosphorylation time interval decreases with \ATPfrac{}
such that the compensation for a stable period does not occur.
(E) As the ATP fraction increases, the number of hexamers or monomers that go through a full cycle
per period increases, such that the amount of KaiA that can be sequestered increases.
(F) Therefore, the time interval per period that all KaiA in the system is sequestered by KaiC,
becomes longer with increasing \ATPfrac{}.
}
\end{figure}

\begin{figure}[t!]
{\centering
\includegraphics[scale=1.0]{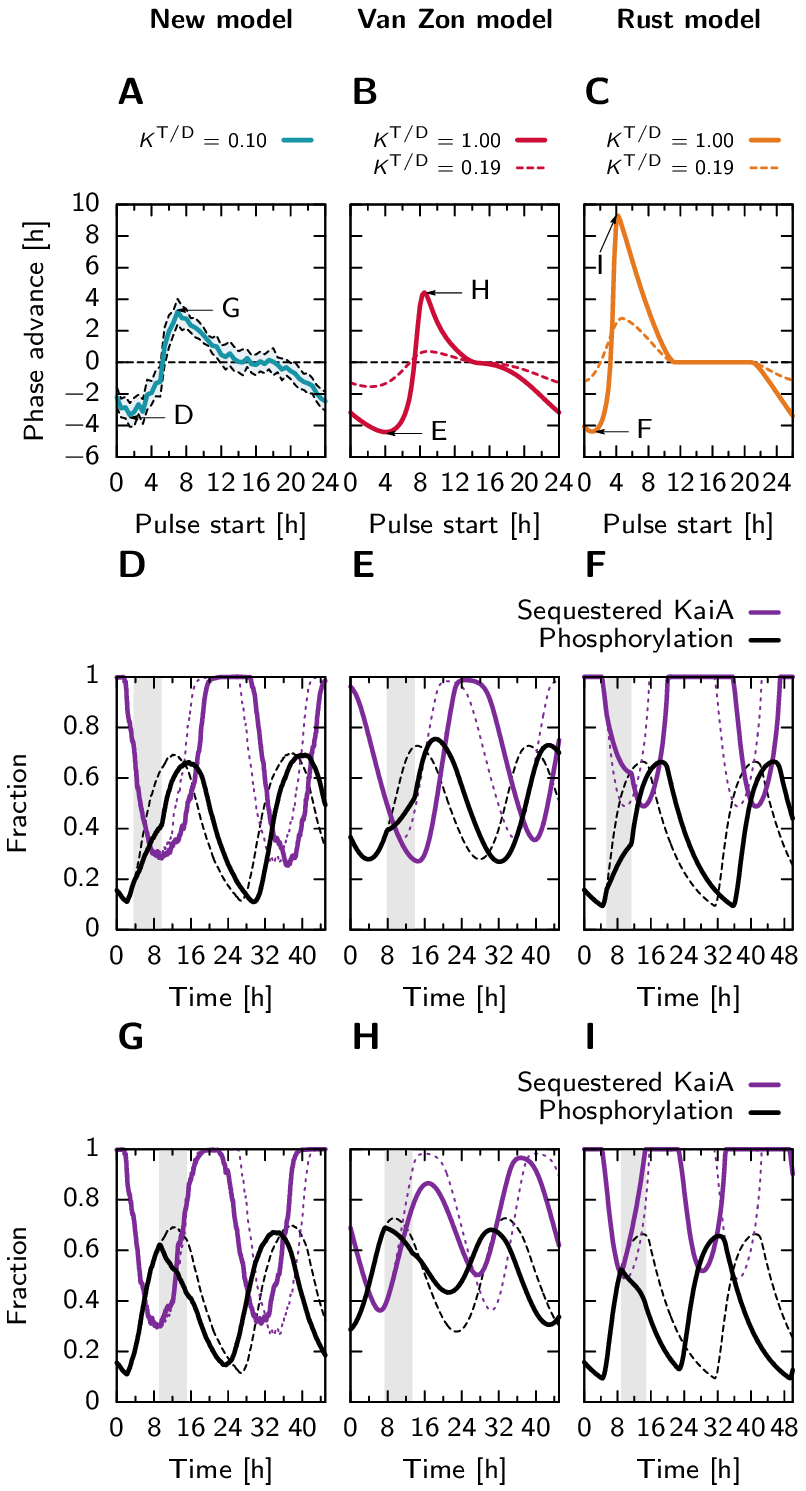}
}
\caption{\flabel{ModelATPSensitivity} 
Sensitivity of the phase of the phosphorylation level, in the Paijmans model (left column), 
Van Zon model (middle column) and the Rust model (right column), 
to a 6 hour pulse of lowering of the ATP fraction to 40\%.
Panels A, B and C show the phase advance in hours,
due to the pulse, starting at the time indicated on the x-axis,
measured from the minimum in the phosphorylation level.
The shaded region in panel A shows the standard deviation of 10 independent runs 
of the Paijmans model, which, contrary to the other two models, is stochastic.
For the Van Zon and Rust models, 
we show phase response curves for a scenario with an equal relative affinity for ATP and ADP, 
$\rmlabels{K}{ATP/ADP}{}=1.0$ (solid line),
and a version with a lower affinity for ADP, $\rmlabels{K}{ATP/ADP}{}=0.19$ (dashed line).
Arrows indicate extrema in the phase response curves, 
for which we show the corresponding time traces in the panels with the adjacent label.
Panels D,E and F show the effect of a pulse (shaded region) when the phase delay is the largest,
on the phosphorylation level (solid black line) and the fraction sequestered KaiA (solid purple line).
Dashed lines show the development in the case no pulse is given.
Panels G,H and I show the effect of the pulse when the phase advance is the largest.
Results for the Paijmans model were obtained using kinetic Monte Carlo,
and ODE's for the Van Zon and Rust models.
}
\end{figure}

\begin{figure}[t!]
{\centering
\includegraphics[scale=1.0]{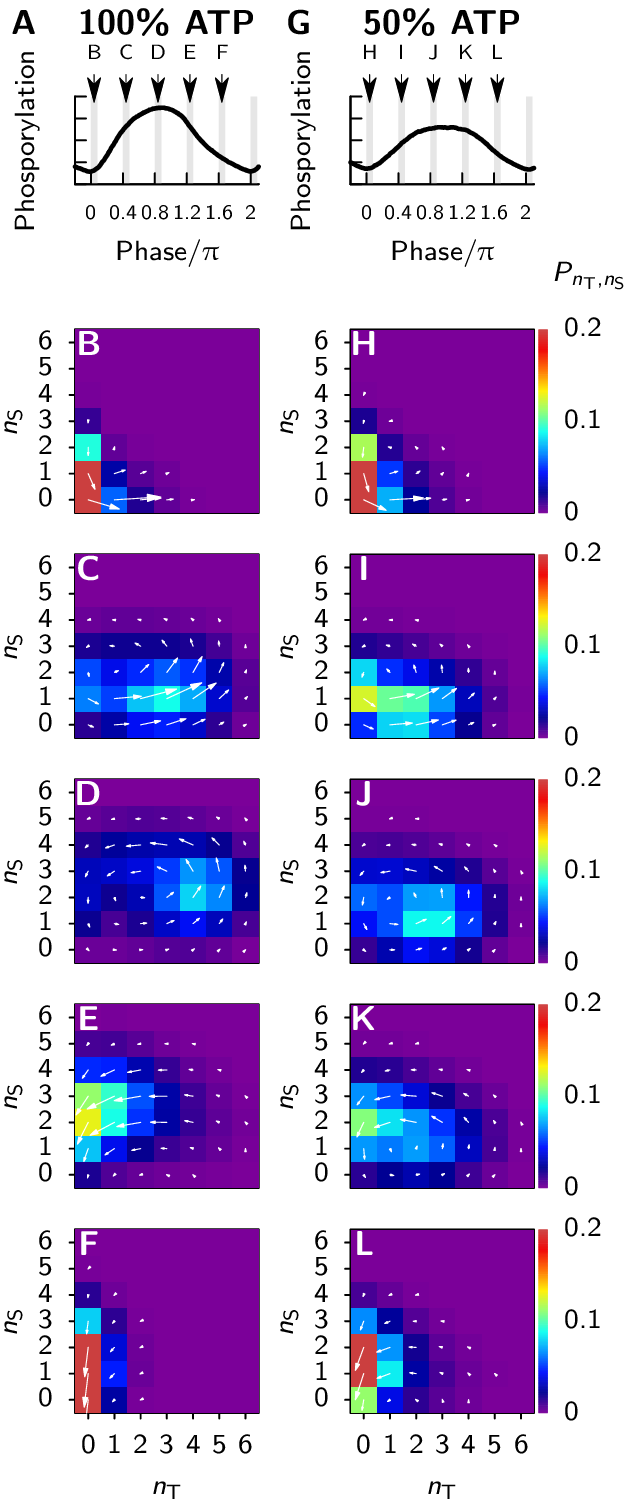}
}
\caption{\flabel{PhaseFlux_ATPcompare} 
In the Paijmans model, at a lower bulk ATP fraction, individual hexamers make a smaller cycle 
through phosphorylation state space.
We track the number of phosphorylated threonine sites, \rmlabels{n}{}{T},
and serine sites, \rmlabels{n}{}{S}, in each individual hexamer for one simulated
hour at different phases of the oscillation.
In panels A and G, 
the shaded regions indicate at which phase of the oscillation hexamers were probed 
and in which panel the result is shown.
Panels B-F (\ATPfrac{}=100\%, left column) and H-L (\ATPfrac{}=50\%, right column) 
show histograms of the probability of finding a hexamer in a certain phosphorylation state,
indicated by the color bar to the right of the right column.
Arrows are proportional to the flux through the state the arrow originates from.
Comparing states near the peak of the phosphorylation level, panels D and J, 
hexamers at \ATPfrac{}=50\% go through a smaller cycle
compared to the situation at 100\% ATP.
Specifically, the majority of hexamers go through \rmlabels{n}{}{T}=4 or 5
at \ATPfrac{}=100\%, while at 50\% hexamers only reach 
\rmlabels{n}{}{T}=2 or 3.
Furthermore, the ensemble is less synchronized near the trough of the 
phosphorylation levels, comparing panels F and L.
Results shown are averaged over 400 consecutive oscillation cycles.
}
\end{figure}

\begin{figure}[t!]
{\centering
\includegraphics[scale=1.0]{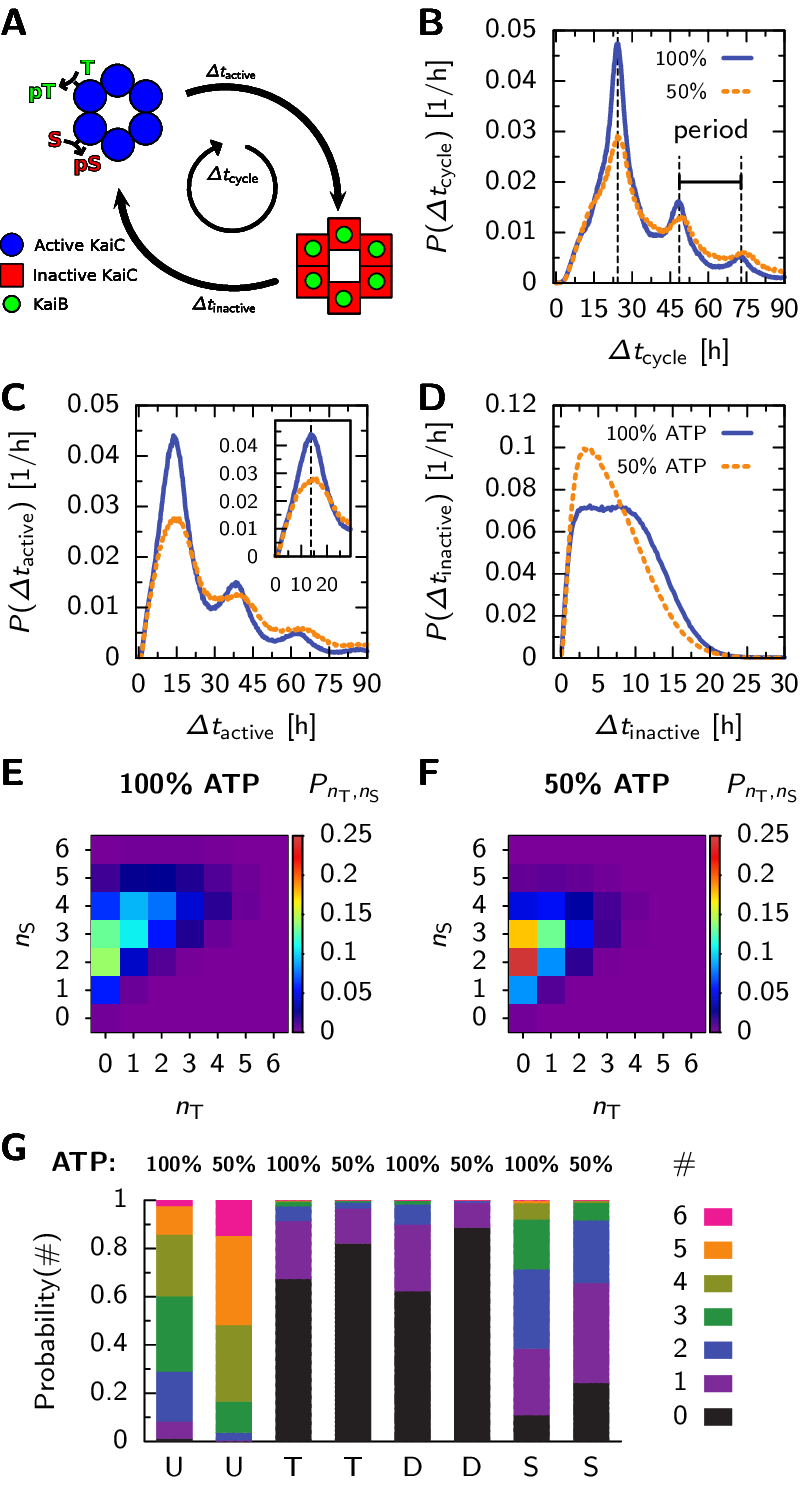} 
\par
}
\caption{\flabel{MicroscopicDynamics} 
Microscopic dynamics of the new model reveals input compensation at the level of individual hexamers.
(A) For each hexamer we measure, 
1) $\Delta\rmlabels{t}{}{active}$: The time between the first phosphorylation event,
and having six KaiB bound to the hexamer for the first time. 
2) $\Delta\rmlabels{t}{}{inactive}$: The time between having 6 KaiB bound and, 
after dephosphorylation and switching back to the active state, 
the first phosphorylation event.
The time for completing a full cycle is: 
$\Delta\rmlabels{t}{}{cycle}=\Delta\rmlabels{t}{}{active}+\Delta\rmlabels{t}{}{inactive}$.
}
\flushright \emph{(continued next page)}
\hrule 
\end{figure}
\begin{figure}[t!]
\ContinuedFloat
\caption{(\emph{continuing from previous page}) 
B, C and D show histograms of these time intervals,
comparing situations with 100\% ATP (blue solid lines)
and 50\% ATP in the bulk (orange dashed lines).
(B) Distribution of times for completing a cycle. 
Peaks are at multiples of the oscillator's period of 24.3 hrs, 
as indicated by the vertical dotted lines (period at 100\% ATP).
Peaks at $\Delta\rmlabels{t}{}{cycle}>24.3$hrs 
show hexamers that could not complete 
a full cycle during one period of the oscillation.
(C) Distribution of times for phosphorylation and KaiB binding.
Event though phosphorylation rates are lower at \ATPfrac{}=50\% as compared to 100\%,
the modes of their distributions are remarkably similar,
as emphasized by the inset which zooms in on the first peak.
(D) Distribution of times for dephosphorylation and waiting for a new round of phosphorylation.
The bigger shoulder at \ATPfrac{}=100\% is a manifestation of the longer
time during which all KaiA is sequestered at this ATP level.
At the end of their cycle, 
hexamers have to wait longer before KaiA returns to solution and phosphorylation starts again.
(E,F) Histograms of the number of phosphorylated threonine sites, \rmlabels{n}{}{T},
and serine sites, \rmlabels{n}{}{S}, in individual hexamers, at the moment
when they switch to the inactive state, at \ATPfrac{}=100\% (E) and 50\% (F).
(G) Histograms of the number (different colors, defined on the right side) 
of U,T,D and S phosphorylated monomers inside a hexamer, 
at the moment when all KaiA in the system is sequestered.
At \ATPfrac{}=100\% more monomers are doubly phosphorylated.
}
\smallskip
\hrule 
\end{figure}

\begin{figure}[t!]
{\centering
\includegraphics[scale=1.0]{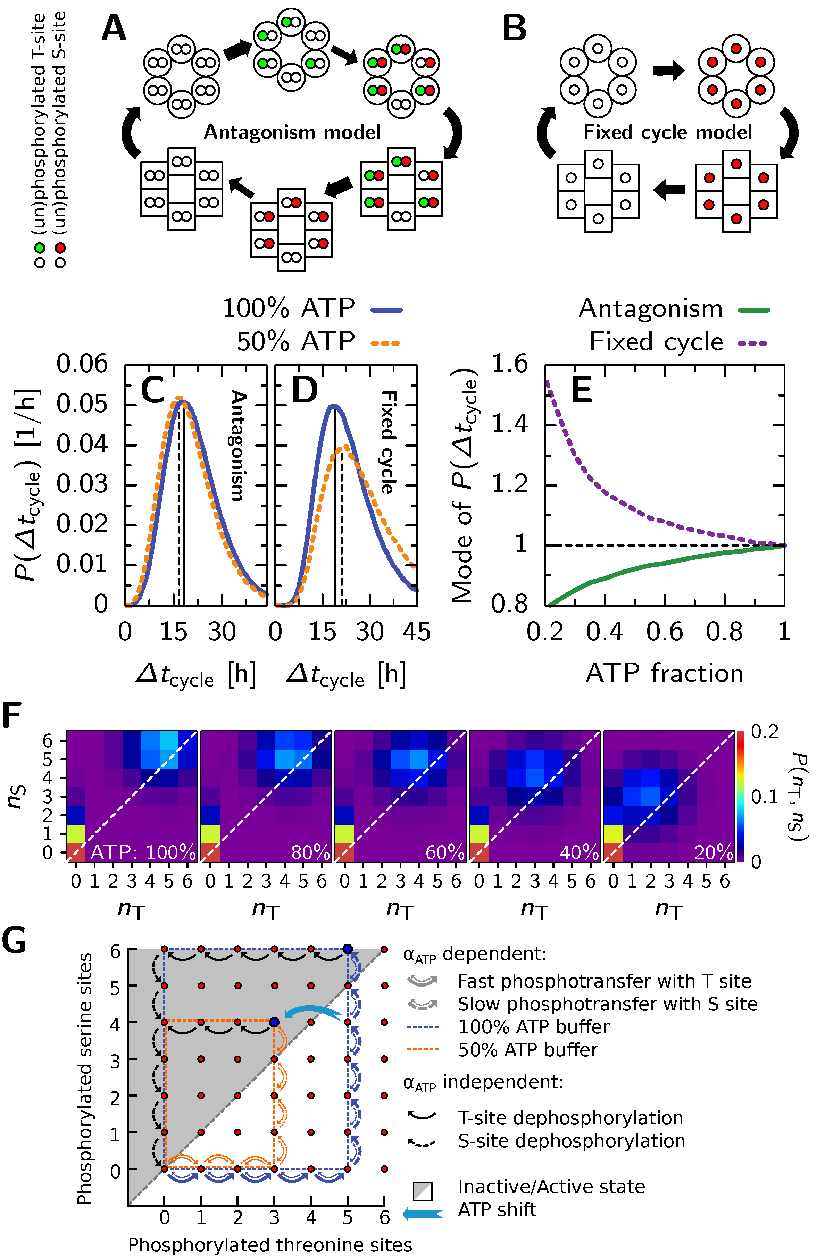}
}
\caption{\flabel{SimpleModel} Antagonistic regulation of the timing of
the conformational switch of KaiC creates input compensation in a single hexamer.
In the antagonism model (A), fast phosphorylation of the threonine sites (green dots)
stabilizes the active conformation (round monomers),
and slow phosphorylation of the serine sites (red dots)
stabilizes the inactive state (square monomers).
Phosphotransfer rates depend on \ATPfrac{} in the active state, but not in the inactive state.
We compare this model with the fixed cycle model (B),
where the phosphorylation of a monomer's single phosphorylation site (red dot)
stabilizes the inactive conformation.
(C,D) Distribution of cycle times, $\Delta\rmlabels{t}{}{cycle}$,
defined as the time interval between two consecutive switches to the active state,
for the antagonism model (C) and the fixed cycle model (D)
at 100\% (blue solid line) and 50\% (orange dashed) ATP buffer.
(E) The mode of the distribution of cycle times for different ATP fractions in the buffer,
for the antagonism (green solid line) and fixed cycle (purple dashed) model.
(F) Distribution in phosphorylation state-space, $P(\npT,\npS)$,
of where the hexamer switches to the inactive state (top-right, along diagonal)
and the active state (lower left corner), at different bulk ATP fractions.
As \ATPfrac{} decreases, the hexamer flips to the inactive state at lower
phosphorylation levels.
(G) Cartoon shows typical cycles through phosphorylation state-space
for a hexamer in the antagonism model in a 100\% (blue trajectory) and 50\% (orange) ATP buffer.
At 100\% ATP, the equilibrium level of phosphorylated threonine sites
in the active state is higher than at 50\% ATP, forcing
the hexamer through a larger cycle as it can only change conformation
when it's left of the diagonal through the state space (gray area).
This compensates the fact that it progresses through the cycle at a higher rate,
such that the time to complete a full cycle is less affected.
A temporary lowering of the ATP level from 100 to 50\%,
can induce a phase advance as the phosphorylation level of the threonine sites 
adjust to a lower equilibrium level allowing the hexamer to switch to the
inactive state, as indicated by the light blue arrow.
}
\end{figure}
\FloatBarrier

\newpage

\end{document}